\documentclass[a4paper,11pt]{article}
\pdfoutput=1 

\usepackage{jheppub} 

\usepackage[T1]{fontenc} 

\usepackage{multirow}
 
\usepackage{rotating}
\usepackage{longtable}
\usepackage{dcolumn}  

\usepackage{lineno}
\include{patch_amsmath}

\usepackage{booktabs}
\usepackage{orcidlink}

\usepackage[tableposition=above]{caption}

\renewcommand{\arraystretch}{1.1}

\newcommand{\cm}{\mathrm{cm}}

\newcommand{\mev}{\mathrm{MeV}}
\newcommand{\mevc}{\mathrm{MeV}/c}
\newcommand{\mevm}{\mathrm{MeV}/c^2}
\newcommand{\gev}{\mathrm{GeV}}
\newcommand{\gevc}{\mathrm{GeV}/c}
\newcommand{\gevm}{\mathrm{GeV}/c^2}

\newcommand{\ee}{e^+e^-}
\newcommand{\uu}{\mu^+\mu^-}
\newcommand{\pp}{\pi^+\pi^-}

\newcommand{\U}{\Upsilon}
\newcommand{\Ufo}{\Upsilon(4S)}
\newcommand{\Uf}{\Upsilon(5S)}

\newcommand{\et}{\eta_b(1S)}

\newcommand{\jp}{J/\psi}

\newcommand{\ks}{K^0_S}

\newcommand{\bp}{B^+}
\newcommand{\bn}{B^0}
\newcommand{\kp}{K^+}
\newcommand{\km}{K^-}
\newcommand{\pip}{\pi^{+}}

\newcommand{\pim}{\pi^{-}}
\newcommand{\mbc}{M_{\rm bc}}

\newcommand{\DE}{\Delta E}
\newcommand{\fb}{\mathrm{fb}^{-1}}

\newcommand{\lik}{\mathcal{L}}

\newcommand{\ecm}{E_{\rm cm}}

\newcommand{\deb}{\Delta E_{\rm BaBar}}
\newcommand{\nr}{n}
\newcommand{\sh}{s}
\newcommand{\ff}{\phi}

\newcommand{\ah}{a_\mathrm{h}}
\newcommand{\bb}{B\bar{B}}
\newcommand{\bbst}{B\bar{B}^*}
\newcommand{\bstbst}{B^*\bar{B}^*}
\newcommand{\bball}{B^{(*)}\bar{B}^{(*)}}

\newcommand{\opdi}{(1+\delta_\mathrm{ISR})}
\newcommand{\PB}{{\cal P}_{B}}

\newcommand{\bs}{B_s^0}

\title{\boldmath Measurement of the energy dependence of the
  $e^+e^-\to B\bar{B}$, ${B}\bar{B}{}^*$, and ${B}^*\bar{B}{}^*$ 
  cross sections at Belle~II}

\collaboration{The Belle II collaboration}
  \author{I.~Adachi\,\orcidlink{0000-0003-2287-0173},} 
  \author{L.~Aggarwal\,\orcidlink{0000-0002-0909-7537},} 
  \author{H.~Ahmed\,\orcidlink{0000-0003-3976-7498},} 
  \author{H.~Aihara\,\orcidlink{0000-0002-1907-5964},} 
  \author{N.~Akopov\,\orcidlink{0000-0002-4425-2096},} 
  \author{A.~Aloisio\,\orcidlink{0000-0002-3883-6693},} 
  \author{N.~Althubiti\,\orcidlink{0000-0003-1513-0409},} 
  \author{N.~Anh~Ky\,\orcidlink{0000-0003-0471-197X},} 
  \author{D.~M.~Asner\,\orcidlink{0000-0002-1586-5790},} 
  \author{H.~Atmacan\,\orcidlink{0000-0003-2435-501X},} 
% ...

\abstract{ We report measurements of the $e^+e^-\to{B}\bar{B}$,
  ${B}\bar{B}^*$, and ${B}^*\bar{B}^*$ cross sections at four
  energies, $10653$, $10701$, $10746$ and $10805\,\mev$, using data
  collected by the Belle~II experiment. We reconstruct one $B$ meson
  in a large number of hadronic final states and use its momentum to
  identify the production process. In the first $2-5\,\mev$ above
  ${B}^*\bar{B}^*$ threshold, the $e^+e^-\to{B}^*\bar{B}^*$ cross
  section increases rapidly. This may indicate the presence of a pole
  close to the threshold. }

\keywords{e+e- Experiments, Quarkonium, Spectroscopy}

\begin{document}
\maketitle
\flushbottom

\section{Introduction}

All known hadrons that contain a heavy $b\bar{b}$ quark pair and have
mass above the open bottom ($\bb$) threshold, $\Ufo$, $\U(10860)$, and
$\U(11020)$, exhibit anomalous properties~\cite{Bondar:2016hva}. In
particular, the $\pp$ transitions to lower bottomonium levels are
strongly enhanced compared to similar transitions from below-threshold
states, and the $\eta$ transitions are not strongly suppressed
compared to the $\pp$ transitions; the latter property violates
heavy-quark spin symmetry. These unexpected properties could be
explained if the hadrons have an exotic admixture: for example, in
addition to $b\bar{b}$, they may also contain multiquark
$b\bar{b}q\bar{q}$ or hybrid $b\bar{b}g$ components, where $q$ and $g$
represent a valence light quark and valence gluon,
respectively~\cite{Meng:2007tk,Simonov:2008ci,Kaiser:2002bm,Voloshin:2012dk}.

In 2019, the Belle experiment observed a new structure, $\U(10753)$,
in the energy dependence of the $\ee\to\U(nS)\pp$ ($n=1,2,3$) cross
sections~\cite{Belle:2019cbt}. While the global significance of the
observation is 5.2 standard deviations, an independent confirmation
would be important. There is a dip in the total $b\bar{b}$ cross
section at the position of the new state, which could be due to 
destructive interference with other contributions~\cite{Dong:2020tdw}.

Recently, the Belle experiment measured the energy dependence of the
$\ee\to\bb$, $\bbst$, and $\bstbst$ cross
sections~\cite{Belle:2021lzm}. These open-flavor final states are
expected to be the dominant decay channels for $b\bar{b}$ hadrons and
constitute the main contribution to the total $b\bar{b}$ cross
section. In fact, their measurement enabled a combined analysis of all
available energy-scan results~\cite{Husken:2022yik}.
The following cross sections were considered: $\ee\to\bb$, $\bbst$,
$\bstbst$, ${B}_s^{(*)}\bar{B}_s^{(*)}$, $\U(nS)\pp$ ($n=1,2,3$),
$h_b(mP)\pp$ ($m=1,2$), and the total $b\bar{b}$ cross section.
The coupled-channel approach was used, which takes into account
rescattering between different channels.
The results of the combined analysis were the pole positions of the
$\U$ resonances and the energy dependence of the scattering
amplitudes. The combined analysis provided further confirmation of the
$\U(10753)$ state. However, its pole position has large
uncertainty. In addition, the scattering amplitudes have large
uncertainties in the $\U(10753)$ region where the spacing between the
Belle scan points is large, about $50\,\mev$.

In order to improve understanding of the $\U(10753)$ energy region,
the SuperKEKB collider performed an energy scan in November 2021. Four
data samples have been collected by the Belle~II experiment;
corresponding center-of-mass (c.m.)\ energies and integrated
luminosities are shown in Table~\ref{tab:lum_euu}.
\begin{table}[htbp]
\caption{ Center-of-mass energy and integrated luminosity of the scan
  data samples. The uncertainties in the c.m.\ energy shown here are
  uncorrelated point-to-point; the correlated uncertainty is
  $0.5\,\mev$. The uncorrelated uncertainty in the luminosity is
  negligibly small; the correlated uncertainty is $0.6\%$. }
\label{tab:lum_euu}
\centering
\begin{tabular}{@{}ccc@{}}
  \toprule
  Point\# & $\ecm\;(\mev)$ & $L\;(\fb)$ \\
  \midrule
  1 & $10804.50\pm0.70$ & 4.690 \\
  2 & $10746.30\pm0.48$ & 9.818 \\
  3 & $10700.90\pm0.63$ & 1.633 \\
  4 & $10653.30\pm1.14$ & 3.521 \\
  \bottomrule
\end{tabular}
\end{table}
The energies have been chosen to fill the gaps between the Belle scan
points. The integrated luminosities are larger than those at Belle,
which are approximately $1\,\fb$ per point. The largest sample is
collected at the expected $\U(10753)$ peak position so that
$\U(10753)$ decays can be studied. Using these data, Belle~II measured
the $\ee\to\U(nS)\pp$ ($n=1,2,3$) cross sections, confirming the
$\U(10753)$ state with high significance~\cite{Belle-II:2024mjm};
observed a strong enhancement of the $\ee\to\chi_{bJ}(1P)\,\omega$
($J=1,2$) cross sections in the $\U(10753)$ region, establishing the
$\U(10753)\to\chi_{b1}(1P)\,\omega$ decay
channel~\cite{Belle-II:2022xdi}; and set stringent upper limits on the
$\ee\to\et\omega$ and $\ee\to\chi_{b0}(1P)\omega$ cross sections near
the $\U(10753)$ peak~\cite{Belle-II:2023twj}.

In this paper, we report measurements of the $\ee\to\bb$, $\bbst$, and
$\bstbst$ cross sections using the scan data collected by
Belle~II. Our analysis closely follows that of
Belle~\cite{Belle:2021lzm}. We perform a full reconstruction of one
$B$ meson in hadronic channels, and then identify the $\bb$, $\bbst$,
and $\bstbst$ signals using the $\mbc$ distribution,
\begin{equation}
  \mbc=\sqrt{(\ecm/2)^2-p_B^2},
\end{equation}
where $\ecm$ is the c.m.\ energy of the colliding beams and $p_B$ is
the $B$-candidate momentum in the c.m.\ frame.
In all equations in this paper, we use natural units $c=1$.
The $\mbc$ variable is essentially a transformation of the $B$-meson
momentum that provides a simpler parametrization of the background
near the kinematic end-point.
Photons from $B^*\to{B}\gamma$ decays are not reconstructed.
To reconstruct $B$ mesons in a large number of hadronic final states, 
we apply the multivariate full event interpretation (FEI)
algorithm~\cite{Keck:2018lcd}.
We use an FEI configuration optimized for energy-scan
analyses~\cite{Belle:2021lzm}. 
The optimization primarily concerns FEI inputs, which are chosen to
make the reconstruction efficiency independent of energy. The $\DE$
variable is not included in the training and a sideband in the
$(\mbc,\;\DE)$ plane is used to constrain backgrounds. The $\DE$ 
variable is defined as
\begin{equation}
\DE = E_B - \ecm/2,
\end{equation}
where $E_B$ is the $B$-candidate energy measured in the
c.m.\ frame. 
The absolute value of the reconstruction efficiency is determined
using $\Ufo$ data. In the $\mbc$ fits, the signals are described using
a function developed in Ref.~\cite{Belle:2021lzm} that is calculated
numerically and includes all relevant effects, in particular,
initial-state radiation (ISR) and the energy dependence of the cross
sections. The latter is determined by fitting both the cross sections
measured in this analysis and the results of the Belle
measurement~\cite{Belle:2021lzm}. To obtain self-consistent results,
an iterative procedure is used.

The paper is organized as follows. We describe the Belle~II detector
and data sets in Section~\ref{sec:data_sets}. The selection of events
using the FEI is presented in Section~\ref{sec:selection}. We discuss the
calibration of the Monte Carlo (MC) simulation, optimization of the
final selection requirements, and determination of the dependence of
the FEI efficiency on c.m.\ energy in
Section~\ref{sec:calib_optim}. Section~\ref{sec:abs_eff} is devoted to
the measurement of the absolute value of the FEI efficiency using
$\Ufo$ data. The fits to $\mbc$ at the scan energies, the fit to the
energy dependence of the $\bball$ cross sections, and the evaluation
of the systematic uncertainties are described in
Section~\ref{sec:scan}. Discussion of the results and summary are in
Section~\ref{sec:discussion}.

\section{Belle II detector and data sets}
\label{sec:data_sets}

The analysis is based on the data collected by the Belle~II
detector~\cite{Belle-II:2010dht} operating at the SuperKEKB
asymmetric-energy $\ee$ collider~\cite{Akai:2018mbz} at KEK.

The Belle II detector is a cylindrical large-solid-angle magnetic
spectrometer consisting of a silicon pixel detector surrounded by a
four-layer double-sided silicon strip detector (SVD) and a 56-layer
central drift chamber (CDC) providing information about charged
particle trajectories (tracks) and vertex positions. Surrounding the
CDC, a time-of-propagation counter (TOP) in the barrel region and an
aerogel-based ring-imaging Cherenkov counter (ARICH) in the endcap
region provide charged-particle identification (PID). Surrounding the
TOP and ARICH, an electromagnetic calorimeter based on CsI(Tl)
crystals provides energy and timing measurements for photons and
electrons. These sub-systems are surrounded by a superconducting
solenoid, providing an axial magnetic field of 1.5~T. An iron flux
return located outside the coil is instrumented with resistive plate
chambers and plastic scintillators to detect $K^0_L$ mesons and to
identify muons (KLM). More details about the detector are given in
Ref.~\cite{Belle-II:2010dht}.

The z axis of the Belle II detector is defined as the symmetry axis of
the solenoid, and the positive direction is approximately given by the
electron-beam direction. The polar angle $\theta$, as well as the
longitudinal and the transverse directions, are defined with respect
to the z axis.

The data analysis strategy is tested on simulated event samples.
Events containing $B$ mesons are generated using the EvtGen
package~\cite{Lange:2001uf}. Continuum $\ee\to q\bar{q}$ background
processes, where $q=u$, $d$, $s$, $c$, are generated with
{\small KKMC}~\cite{Jadach:1999vf} and
{\small PYTHIA}\,8~\cite{Sjostrand:2014zea}.
Final-state radiation of photons from stable charged particles is
simulated with PHOTOS~\cite{Barberio:1990ms}.
The detector response and $K^0_S$ decays are simulated using
Geant4~\cite{GEANT4:2002zbu}. Both collision data and simulated
samples are processed using the Belle~II software~\cite{Kuhr:2018lps}.

We use the Belle II energy-scan data, consisting of the four samples
with energies and luminosities shown in Table~\ref{tab:lum_euu}. To
determine the FEI efficiency, we use $\Ufo$ data taken immediately
before and after the energy scan period; the combined integrated
luminosity of this $\Ufo$ data sample is $35.5\,\fb$.

\section{Event selection}
\label{sec:selection}

In the configuration of the FEI used here
(following~\cite{Belle:2021lzm}), we reconstruct the $B^+$ and $B^0$
mesons in the decay channels $\bar{D}^{(*)}\pi^+(\pi^+\pi^-)$,
$D_s^{(*)+}\bar{D}^{(*)}$, $\jp\kp(\pim)$, $\jp\ks(\pip)$,
$\jp\ks\pp$, $D^{(*)-}\pi^+\pi^+$, and $D^{*-}K^+K^-\pi^+$, where
$\bar{D}$ denotes the $\bar{D}^0$ and $D^-$
mesons.\footnote{Throughout this paper, charge conjugated channels are
  always included.}
The $D^0$, $D^+$, and $D_s^+$ mesons are reconstructed in final
states with $K^\pm$, $\ks$, $\pi^\pm$, up to one $\pi^0$, and
multiplicity up to five.
The list of channels that are used for reconstruction of $B$ and $D$
mesons is presented in Appendix~\ref{sec:fei_b_d_chan}.
We reconstruct $D^*$ mesons in the $D\pi$ and $D\gamma$ channels;
$\jp$ are reconstructed in the $\uu$ and $\ee$ final states.

We perform a loose selection of the final-state particles and decays,
and subsequently use multivariate analysis for the final selection.
We select tracks that originate from the vicinity of the interaction
point (IP) by imposing $dr<0.5\,\cm$ and $dz<3\,\cm$, where $dr$ and
$dz$ are transverse and longitudinal distances between the track and
the IP.
The PID is based on the ionization energy-loss measurement in the CDC
and responses of the TOP, ARICH, ECL, and KLM. Information from these
subdetectors is combined into a likelihood $\lik_h$ for a given
hypothesis $h$~\cite{Belle-II:2018jsg}, and the ratio
$R_h=\lik_h/(\lik_{e^+}+\lik_{\mu^+}+\lik_{\pi^+}+\lik_{K^+}+\lik_{p}+\lik_{d})$
is used in the selection. In the initial selection, we apply the PID
requirement only for kaon candidates, $R_{K^+}>0.1$.  The efficiency
of this requirement is 86\% and the probability to misidentify a pion
as a kaon is about 7\%.
We require photons to have energies greater than $100$, $90$, and
$160\,\mev$ in the forward endcap ($12.4^\circ<\theta<31.4^\circ$),
barrel ($32.2^\circ<\theta<128.7^\circ$), and backward endcap
($130.7^\circ<\theta<155.1^\circ$) regions of the ECL, respectively,
as the backgrounds in these regions are different.
For the $\ks$, $\pi^0$, $D$, and $\jp$ candidates, we apply a loose
mass-range requirement that corresponds to about $\pm5$ units of mass
resolution. For the $D^*$ candidates, we use the mass difference
$M(D^*)-M(D)$.
To improve momentum resolution, we apply a mass-constrained fit to
$\pi^0$, $\jp$, and $D^*$ candidates; a mass-vertex-constrained fit to
$D$ and $D_s^+$; and a vertex-constrained fit to $\ks$ and $B$.

A boosted decision tree~\cite{Keck:2017gsv} is used with the following
discriminating variables for various particle species.
\begin{itemize}
\item For charged pions, kaons, and leptons, we use the laboratory
  momentum, the transverse momentum, and the PID information, which
  consists of likelihood ratios for the $e^+$, $\mu^+$, $\pi^+$,
  $K^+$, and $p$ hypotheses.
\item For photons, we use the laboratory momentum, the polar angle,
  the number of crystals in the energy deposition (cluster), the ratio
  of the energy deposition in a $3\times3$ matrix of crystals to that
  in a $5\times5$ matrix without corner crystals, and the cluster
  timing. These variables are used to suppress hadron showers and beam
  background.
\item
  For $\ks\to\pp$ candidates, we use the invariant mass and laboratory
  momentum of the $\ks$ candidate, the distance between the IP and the
  $\ks$ vertex, the cosine of the angle between the $\ks$ momentum and
  the direction from the IP to the $\ks$ vertex, the distance between
  the $\pip$ and $\pim$ tracks along the beam direction at the $\ks$
  vertex, the numbers of SVD and CDC measurement points (hits) of
  $\pip$ and $\pim$, and the decay angle (the angle between the $\pip$
  momentum measured in the $\ks$ rest frame and the $\ks$ boost
  direction from the laboratory frame).
\item
  For $\pi^0\to\gamma\gamma$ candidates, we use the invariant mass,
  the laboratory momentum of the $\pi^0$ candidate, and its decay angle.
\item
  For $D$ meson candidates, we use the invariant mass and p-value of
  the mass-vertex constrained fit. In three-body decays, we include
  invariant masses of intermediate $\rho\,(\to\pi\pi)$,
  $K^*(\to{K}\pi)$, and $\phi(\to\kp\km)$ resonance candidates.
\item
  For $\jp$ and $D^*$ candidates we use invariant masses.
\item
  For $B$ meson candidates, we use the p-value of the
  vertex-constrained fit. If there is a $D$ meson in the decay, we
  include the distance between the $B$ and $D$ vertices, this distance
  divided by its uncertainty, and the cosine of the angle between the
  $D$ momentum and the direction from the $B$ to the $D$ vertex. If
  there are several pions or kaons in the decay, we include invariant
  masses of intermediate $\rho$, $K^*$, and $a_1(\to\pi\pi\pi)$
  resonance candidates.
\item
  To suppress continuum $\ee\to q\bar{q}$ backgrounds, where $q$
  denotes a $u$, $d$, $s$, or $c$ quark, we use the event-shape
  variable $R_2$ (the ratio of the second to zeroth Fox-Wolfram
  moments~\cite{Fox:1978vu}), the angle between the thrust axes of the
  $B$ candidate and the rest of the event~\cite{Belle-II:2018jsg}, and
  two boolean variables indicating the presence of a muon and an
  electron, respectively, in the rest of the event.
  We consider lepton candidates in the c.m.\ momentum ranges
  $1.0<p_\mu<2.6\,\gevc$ and $0.8<p_e<2.6\,\gevc$ where the
  contribution of leptons from semileptonic $B$ decays is
  enhanced. We require that the leptons are well identified with a
  likelihood ratio above 0.9.
  The efficiencies of this requirement are 90\% and 87\% for muons and
  electrons, respectively; the probabilities to misidentify hadrons as
  leptons are at the level of 5\%.
\end{itemize}

For training, we use a simulated $\Ufo$ sample that corresponds to an
integrated luminosity of $100\,\fb$.
The training is performed separately for each final-state particle
species and for each decay of the unstable particles. The training
result, the classifier output, is the probability that a given
candidate is a signal. In addition to the variables listed above, the
training for each decay also uses the signal probabilities of all
direct decay-products.
To realize this, the training is performed in stages: first only final
state particles and $\ks$ are trained, in the next stage $\pi^0$ and
$\jp$, then $D$ mesons, subsequently $D^*$, and finally $B$ mesons.

\begin{figure}[htbp]
\centering
\includegraphics[width=0.45\linewidth]{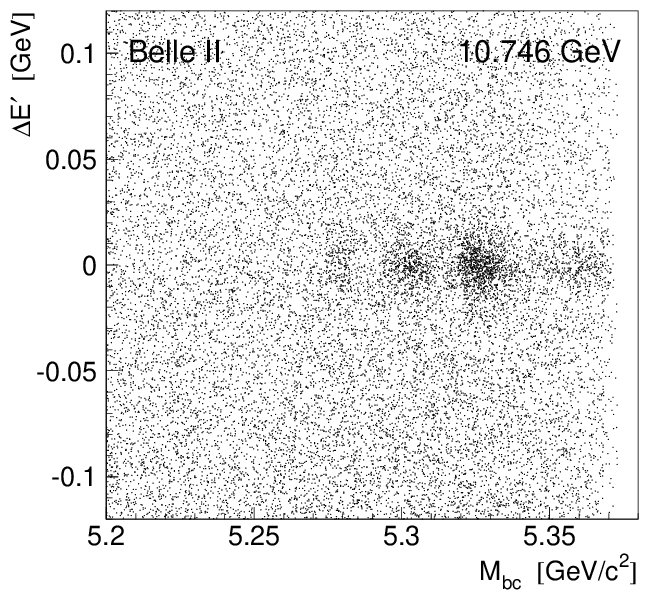}
\caption{ Distributions of $\DE'$ vs.\ $\mbc$ in the
  $\ecm=10.746\,\gev$ data sample. }
\label{de_vs_mbc_10751_210922}
\end{figure}

In the case of multiple $B^+$ ($B^0$) candidates, we select the one
that has the highest signal probability, $\PB$.
Figure~\ref{de_vs_mbc_10751_210922} shows the distribution of $\DE'$
versus $\mbc$ for combined $B^+$ and $B^0$ candidates in the
$\ecm=10.75\,\gev$ data sample. The $\DE'$ variable is defined as
\begin{equation}
\DE' = \DE + \mbc - 5.28\,\gev,
\end{equation}
where the value $5.28\,\gev$ approximates the $B$ meson mass.
Clusters of events are clearly observed at $\DE^\prime\approx0$ with
$\mbc\approx 5.28$, $5.305$, and $5.33\,\gevm$, 
indicating the presence of the $\ee\to\bb$, $\bbst$, and $\bstbst$
processes, respectively.

\section{Simulation studies and corrections}
\label{sec:calib_optim}

\subsection{Calibration of simulation}

Not all channels used for $B$-meson reconstruction are well measured.
Therefore, not all relative yields agree between data and
simulation. We introduce weights for simulated events to mitigate this
problem.

To determine the weights, we use the $\DE'$ distributions in the
$\Ufo$ sample. We select $B$ candidates with requirements $\PB>0.16$
and $\mbc>5.27\,\gevm$.
We perform a simultaneous fit to the $\DE'$ distributions in data and
simulation for each $B$-decay channel. The signal in simulation is
described by a sum of two Gaussian functions; the signal in data is
described by the same model, with the addition of a weight factor, a
shift, and a broadening factor, which are all determined by the
fit. The background is described by a second order polynomial.
The resulting weights are close to $1.0$ for two-body decays, while
for some multibody decays they are as low as $0.4$.

To determine the average shift and broadening factor, we combine all
the channels in both data and simulation; for the latter, we apply the
weights determined above. We then perform the fit to the $\DE'$
distributions in the combined samples.
We find that the shift in $\DE'$ is negligibly small,
$(0.07\pm0.07)\,\mev$, while the scale factor for the width is
\begin{equation}
  \ff = 1.086\pm0.012.
  \label{eq:depr_fufa}
\end{equation}

\subsection{Optimization}

To optimize the selection in the $\PB$ and $\DE'$ variables, we use
a simulated $\ecm=10.751\,\gev$ sample with an effective integrated
luminosity of $50\,\fb$. We maximize the figure-of-merit 
$\mathrm{S}/\sqrt{\mathrm{S}+\mathrm{B}}$, where $\mathrm{S}$ is the
number of properly reconstructed signal candidates, and $\mathrm{B}$
is the number of all other candidates satisfying
$5.270<\mbc<5.335\,\gevm$. The optimization is performed iteratively:
we scan the figure-of-merit in one variable, then in the other. The
resulting values are $|\DE'|<18\,\mev$ and $\PB>0.16$.

\begin{figure}[htbp]
\centering
\includegraphics[width=0.45\linewidth]{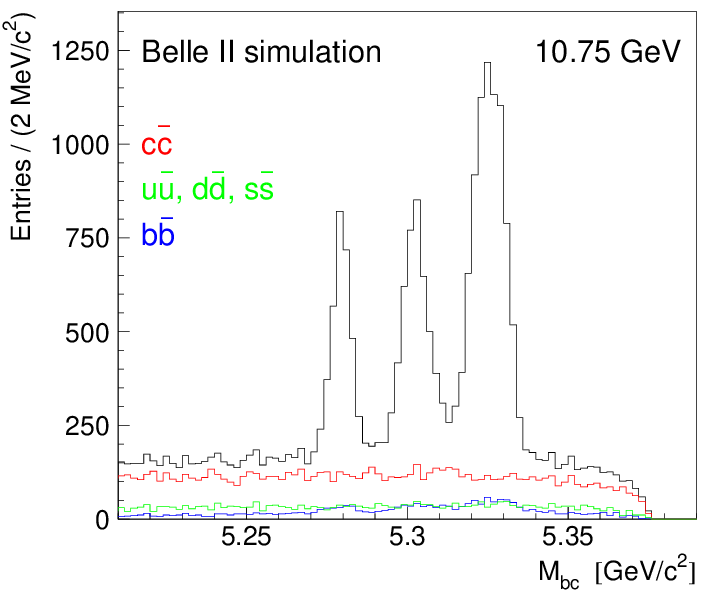}
\caption{ Distribution of $\mbc$ in simulation at
  $\ecm=10.751\,\gev$. The background contributions are also shown:
  $c\bar{c}$ (red), the sum of $u\bar{u}$, $d\bar{d}$, and $s\bar{s}$
  (green), and $b\bar{b}$ (blue). }
\label{mbc_mc_10751_290322}
\end{figure}

Figure~\ref{mbc_mc_10751_290322} shows the $\mbc$ distribution in the
simulation at $\ecm=10.751\,\gev$ after final selection requirements
are applied.
The signal-to-background ratio is relatively high, which is also the
case at the other scan energies. Also shown are the contributions of
the continuum and $b\bar{b}$ backgrounds. Candidates in the $b\bar{b}$
samples that do not correspond to generated signal events are treated
as background. The continuum background, and in particular $c\bar{c}$,
dominates.

\subsection{\boldmath Dependence of the FEI efficiency on $\ecm$}
\label{sec:eff_vs_ecm}

\begin{figure}[htbp]
\centering
\includegraphics[width=0.45\linewidth]{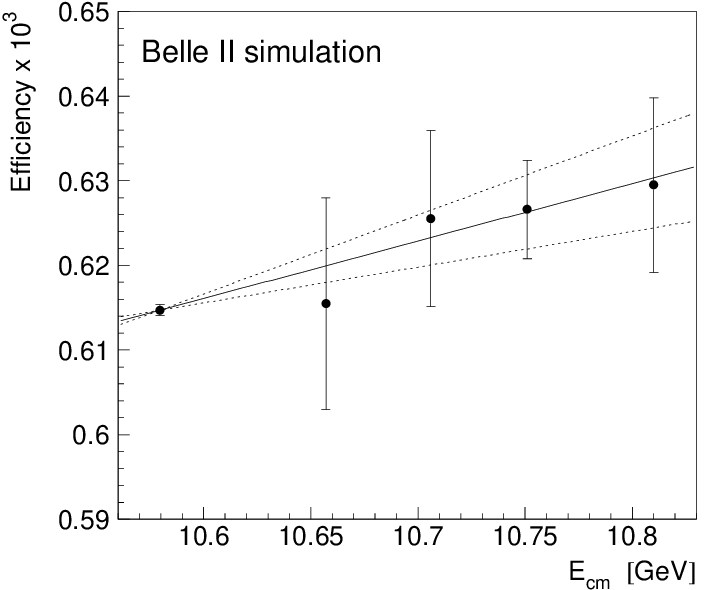}
\caption{ Reconstruction efficiency of the FEI at various
  c.m.\ energies determined using simulation. The solid line is the
  result of the fit to a linear function, and the dashed lines show
  the results when the slope parameter is varied by $\pm1$ standard
  deviation. }
\label{eff_vs_ecm_280322}
\end{figure}

We determine the FEI efficiency as a function of c.m.\ energy using
simulation at the $\Ufo$ and scan energies; the results are shown in
Fig.~\ref{eff_vs_ecm_280322}.
The efficiency increases slightly with energy.
A fit to a linear function gives the following result:
\begin{align}
  \label{eq:eff_vs_ecm}
  \varepsilon = [ & (0.6147\pm0.0007)\; + \\ \nonumber
  & (0.068\pm0.026)\,\gev^{-1}\times(\ecm-10.5796\,\gev)] \times
  10^{-3}.
\end{align}
Here, $10.5796\,\gev$ is the simulated $\Ufo$ energy.
The above fit result corresponds to a $1.032\pm0.012$ ratio between
efficiencies at the $\Uf$ energy ($10.866\,\gev$) and at the $\Ufo$
energy.
This value agrees with the Belle measurement performed using $\Ufo$
and $\Uf$ data, $1.049\pm0.032$~\cite{Belle:2021lzm}.

\section{Absolute value of the FEI efficiency}
\label{sec:abs_eff}

To determine the absolute value of the FEI efficiency we use the
$\Ufo$ data. The efficiency is calculated as
\begin{equation}
  \varepsilon = N/(2\,N_{\bb}),
  \label{eq:abs_eff_eq}
\end{equation}
where $N$ is the number of reconstructed $B$ mesons using the FEI and
$N_{\bb}$ is the total number of $\bb$ events determined by counting
hadronic events and subtracting the continuum contribution:
$N_{\bb} = (38.67 \pm 0.58) \times 10^6$.

The $B$-meson yield $N$ is determined using a fit to the $\mbc$
distribution. We use the fit function that was developed in
Ref.~\cite{Belle:2021lzm}. It is calculated numerically as a sequence
of convolutions and includes the effects of the energy spread of the
colliding beams, initial-state radiation (ISR), the $B$-meson momentum
resolution, and the energy dependence of the production
cross-section. The latter plays an important role if the cross section
changes noticeably over the typical range of the $\ecm$ spread, which
is the case at the $\Ufo$.
In Sections~\ref{sec:mom_resol} and \ref{sec:peaking_bg}, we describe,
respectively, the determination of the momentum resolution function,
and the study of background and broken signal distributions in
simulation. We then present the fit to the $\Ufo$ data, estimate
systematic uncertainties on its results, and calculate the FEI
efficiency (Sections~\ref{sec:fit_y4s_data} to \ref{sec:fei_eff_y4s}).

\subsection{Momentum resolution}
\label{sec:mom_resol}

At the $\Ufo$ energy, the resolution in $\mbc$ is dominated by the
$\ecm$ spread. The $B$-momentum resolution plays a minor role because
it contributes proportionally to the $B$ momentum, which is small at
the $\Ufo$ resonance. However, the effect of the momentum resolution
becomes prominent at scan energies. Thus, we take it into account
consistently for all data samples.

The momentum resolution is parameterized as
\begin{align}
  f(p-p_0) =  \frac{1-r_2-r_3}{\sigma_1}\; & \mathrm{exp}\left[-\frac{(p-p_0-\mu_1)^2}{2\sigma_1^2}\right]\,
  p\,\left\{1-\mathrm{exp}\left[-\frac{2p(p_0+\mu_1)}{\sigma_1^2}\right]\right\} \nonumber \\
  +\; \frac{r_2}{\sigma_2}\; & \mathrm{exp}\left[-\frac{(p-p_0-\mu_2)^2}{2\sigma_2^2}\right]\,
  p\,\left\{1-\mathrm{exp}\left[-\frac{2p(p_0+\mu_2)}{\sigma_2^2}\right]\right\} \nonumber \\
  +\; \frac{r_3}{\sigma_3}\; & \mathrm{exp}\left[-\frac{(p-p_0-\mu_3)^2}{2\sigma_3^2}\right]\,
  p\,\left\{1-\mathrm{exp}\left[-\frac{2p(p_0+\mu_3)}{\sigma_3^2}\right]\right\},
  \label{eq:mom_res}
\end{align}
where $p$ and $p_0$ are the reconstructed and true $B$-meson momenta,
respectively. This function is a sum of three Gaussians with
parameters $\mu_i$ and $\sigma_i$, and weights $r_i$, each multiplied
by an additional factor. The extra factor takes into account the fact
that $p$ is positive definite, and is obtained by considering the
momentum-resolution function in three dimensions and analytically
integrating out all variables other than $p$.
To determine the parameters of the resolution function, we fit the
$p-p_0$ distribution in the simulated $\Ufo$ sample; the results are
presented in Table~\ref{tab:y4s_mom_res} ($f_\mathrm{A}$ column).
\begin{table}[htbp]
  \caption{ Parameters of the momentum resolution functions of
    Eq.~(\ref{eq:mom_res}) for truth-matched candidates
    ($f_\mathrm{A}$) and unmatched candidates in the $\DE'$ signal
    region ($f_\mathrm{B}$) and sideband ($f_\mathrm{C}$). The
    parameters $w$ are the relative weights of the three peaking
    components. }
  \label{tab:y4s_mom_res}
  \centering 
  \begin{tabular}{@{}cccc@{}} \toprule
    & $f_\mathrm{A}$ & $f_\mathrm{B}$ & $f_\mathrm{C}$ \\
    \midrule
    $w$                  & $1$     & $0.048$ & $0.0149$ \\
    $\mu_1$ ($\mevc$)    & $-0.02$  & $0.7$  &  $-21$ \\
    $\sigma_1$ ($\mevc$) & $4.4$   & $14$    &  $115$ \\
    $r_2$                & $0.48$  & $0.48$  & $0$ \\
    $\mu_2$ ($\mevc$)    & $-0.02$  & $-15$    & $-$ \\
    $\sigma_2$ ($\mevc$) & $8.0$   & $95$    & $-$ \\
    $r_3$                & $0.082$ & $0$     & $0$ \\
    $\mu_3$ ($\mevc$)    & $0.03$  & $-$     & $-$ \\
    $\sigma_3$ ($\mevc$) & $17.1$  & $-$     & $-$ \\
    \bottomrule
  \end{tabular}
\end{table}

\subsection{Backgrounds and broken signal}
\label{sec:peaking_bg}

\begin{figure}[htbp]
\centering
\includegraphics[width=0.55\linewidth]{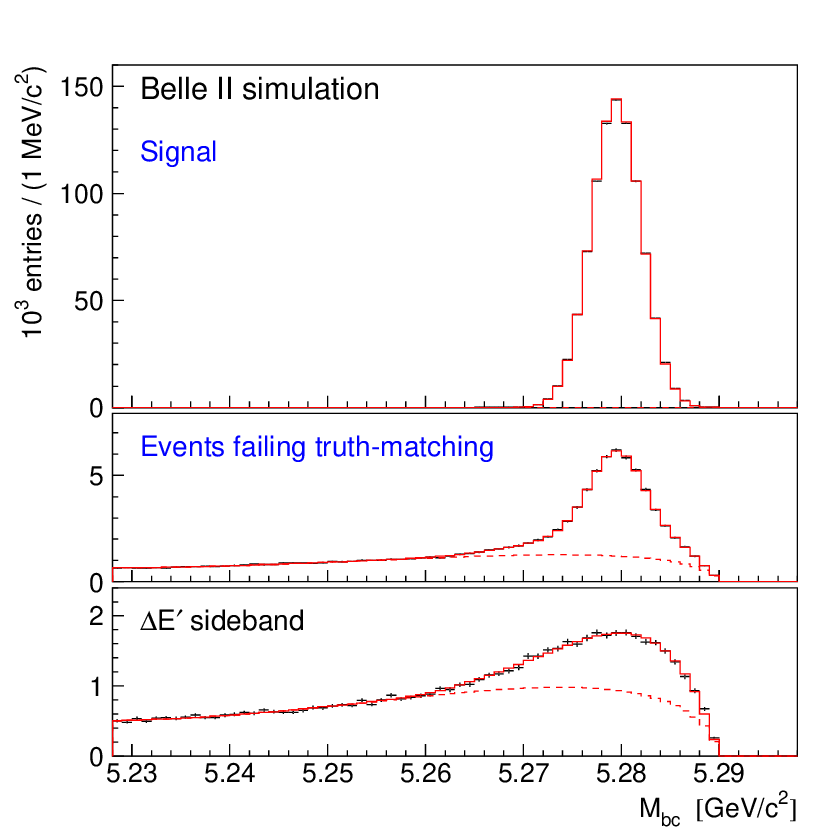}
\caption{ Distributions of $\mbc$ for the simulated $\Ufo$ sample. The
  top and middle panels show truth-matched and non-truth matched
  candidates, respectively, while the bottom panel shows candidates in
  the $\DE'$ sideband. The solid histogram shows the result of the
  simultaneous fit, and the dashed histogram shows the distribution of
  background events. }
\label{mbc_y4s_mc_210322}
\end{figure}

Figure~\ref{mbc_y4s_mc_210322} shows the $\mbc$ distributions for
three categories of candidates in the $\Ufo$ simulation: (1) truth
matched (correctly reconstructed $B$ mesons), (2) non-truth matched,
and (3) candidates in the $\DE'$ sideband. 
The $\DE'$ sideband has the same width as the signal region, while its
center is shifted by $+80\,\mev$.
There is a peaking structure in the distributions of non-truth matched
candidates and candidates in the $\DE'$ sideband.
This structure is due to signal decays in which one of the final state
particles is swapped with a background particle. The magnitude of this
broken signal component is proportional to the signal yield, while its
$\mbc$ distribution is broader due to the incorrectly reconstructed
momentum of the decay products. Thus, it can be described by a $\mbc$
signal function with a poor momentum resolution.

To determine the ratio of the broken signal yield to that of the
signal ($w$) and the parameters of the momentum-resolution function
describing the broken signal, we perform a simultaneous fit to 
the three $\mbc$ distributions shown in Fig.~\ref{mbc_y4s_mc_210322}.
For the truth-matched component, the parameters of the momentum
resolution function are fixed to the values obtained using the
momentum-difference fit described in the previous section. For the
broken signal components, the parameters of the resolution
function are left free.
The simulation does not include the ISR process; we modify the $\mbc$
signal function accordingly. The production cross-section is
energy-independent in the simulation. 

The smooth background is described by a square-root function
multiplied by a third-order Chebyshev polynomial. The parameters of
the polynomial are all determined by the fit. The shape of the smooth
background is the same in the $\DE'$ signal region and in the
sideband, while the normalizations are independent. The fits to $\mbc$
distributions in this paper are binned likelihood fits.

The fit results are shown in Table~\ref{tab:y4s_mom_res} (columns
$f_\mathrm{B}$ and $f_\mathrm{C}$ for the broken signal
components in the $\DE'$ signal region and sideband, respectively). In
the $\DE'$ signal region, two Gaussian functions are sufficient, while
in the sideband, one Gaussian function is sufficient.

\subsection{\boldmath Fit to the $\Ufo$ data}
\label{sec:fit_y4s_data}

To determine the $B$-meson yield in the $\Ufo$ data sample and to find
corrections to the model of the broken signal component, we fit the
$\mbc$ distribution in $\Ufo$ data.
In addition to the steps discussed in the previous section, this
requires the inclusion of ISR and the energy dependence of the cross
section.
Below the $\bbst$ threshold, the $\ee\to\bb$ cross section coincides
with the total $b\bar{b}$ cross section, usually presented in terms of
$R_b$
\begin{equation}
R_b=\frac{\sigma(b\bar{b})}{\sigma_0(\uu)},
\end{equation}
where $\sigma_0(\uu)$ is the Born cross section for $\ee\to\uu$.
We use the most precise measurement of the energy dependence of $R_b$
from Ref.~\cite{Aubert:2008ab}. Since no suitable physics-motivated
model of the $R_b$ shape is available, we use an 11th order Chebyshev
polynomial, as in Ref.~\cite{Belle:2021lzm}. We parameterize
the dressed cross section\footnote{The difference between the dressed
  and Born cross sections is that the former takes into account the
  vacuum polarization effect.} and apply the ISR correction and the
energy-spread correction by performing convolutions with the ISR
radiation kernel~\cite{Kuraev:1985hb} and a Gaussian function with the
standard deviation fixed to the BaBar energy spread, respectively.

The $R_b$-shape measurement~\cite{Aubert:2008ab} is insufficiently
precise to use as a simple input to this procedure: not all curves
that give acceptable fit quality to the $R_b$ scan also satisfactorily
describe the $\mbc$ distribution. Therefore, we perform a simultaneous
fit to the energy dependence of $R_b$ and the $\mbc$ distributions in
the $\DE'$ signal and sideband regions. The inclusion of the $\DE'$
sideband in the fit helps to constrain the smooth background and to
find corrections for the yield and shape of the broken signal
component.

The $R_b$ points have a sizeable energy scale uncertainty of
$1.5\,\mev$~\cite{Aubert:2008ab}. Hence, in Ref.~\cite{Belle:2021lzm}
Belle introduced a common shift in the c.m.\ energy of all the $R_b$
scan points, and determined the value $\deb=(-1.75\pm0.68)\,\mev$ from
the fit. We introduce the $\deb$ shift and fix its value to the Belle
result.

Simulation shows that the scale factor of the momentum resolution is
approximately the same as that of the $\DE'$ signal. Therefore, all
the width parameters of the $f_\mathrm{A}$ component of the resolution
function are multiplied by the scale factor of
Eq.~(\ref{eq:depr_fufa}).
The same scale factor is applied to the narrow Gaussian in the
function $f_\mathrm{B}$ describing the broken signal in the $\DE'$ signal
region.
For the function $f_\mathrm{C}$ describing the broken signal in the
$\DE'$ sideband, we introduce a normalization correction $\nr_3$,
shift $\sh_3$, and width scale factor $\ff_3$. All these parameters
are free in the fit.
Unlike the Belle analysis~\cite{Belle:2021lzm}, we do not use a low
$\DE'$ sideband, as it was found to contain a tail of the signal and a
small contribution from Cabibbo-suppressed decays, such as $DK$ and
$DK\pi\pi$, with the kaon misidentified as a pion. The high $\DE'$
sideband provides more accurate information about the broken signal
component.
The smooth background is parameterized as described in the previous
section for simulation.

\begin{figure}[htbp]
\centering
\includegraphics[width=0.49\linewidth]{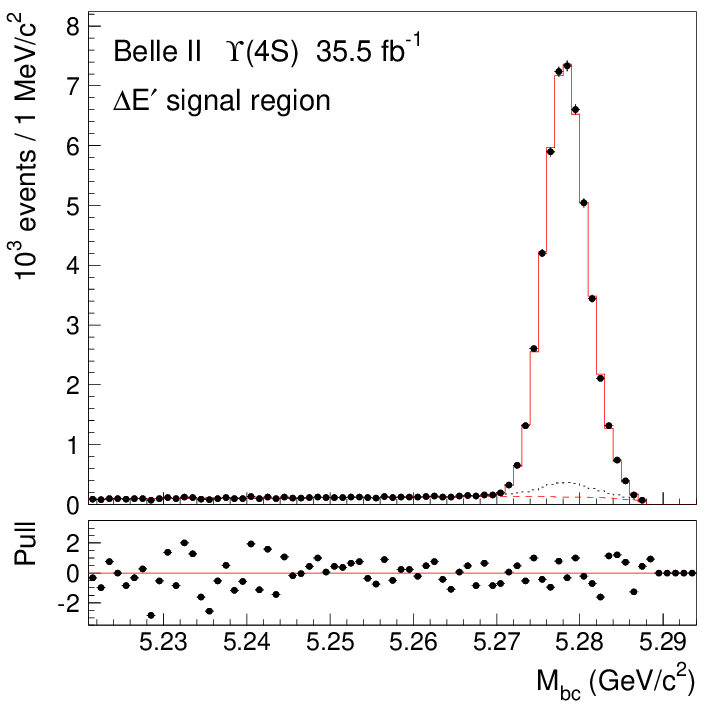}\hfill
\includegraphics[width=0.49\linewidth]{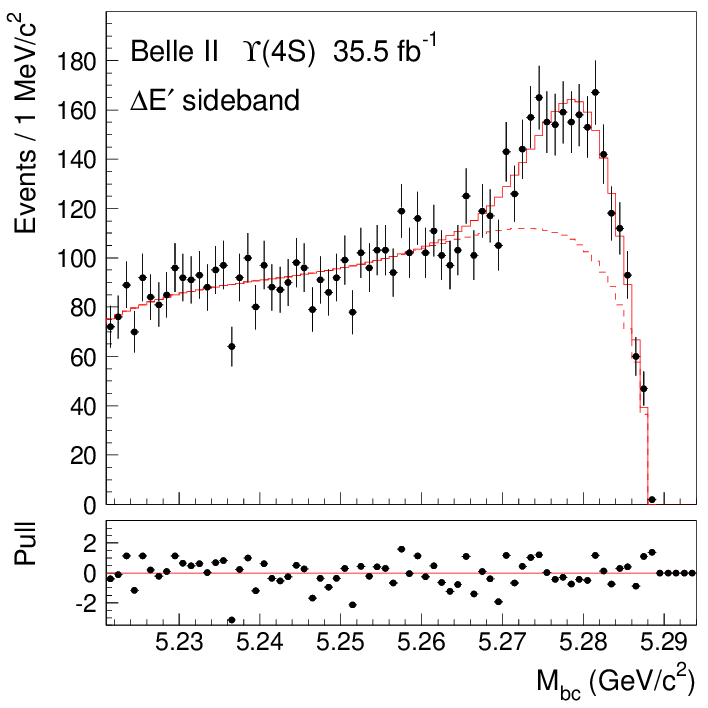}
\caption{ Distributions of $\mbc$ for $\Ufo$ data. The left and right
  panels correspond to the $\DE'$ signal and sideband regions. The
  solid histogram shows the result of the simultaneous fit to these
  distributions and the cross-section energy dependence
  (Fig.~\ref{xsec_fit_rel_xsec_050422}). The red dashed histogram
  shows the background, and the black dotted histogram shows the sum
  of the background and the broken signal (in the right panel, this
  coincides with the total fit, and is thus not visible). The bottom
  panels show pulls (deviations of the data points from the fit
  function divided by the uncertainties on the data).  }
\label{mbc_fit_rel_ecm_240322}
\end{figure}

\begin{figure}[htbp]
\centering
\includegraphics[width=0.57\linewidth]{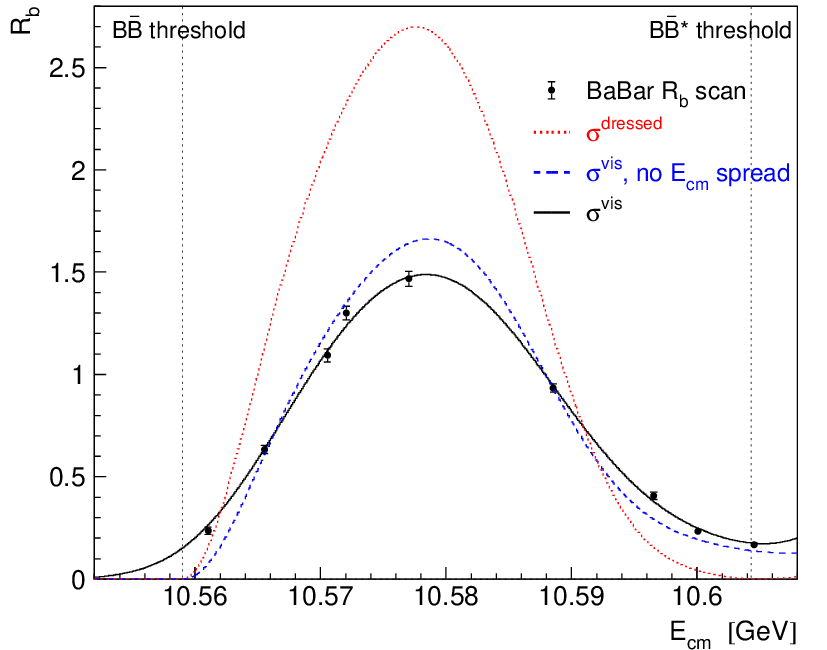}
\caption{ Energy dependence of the total $b\bar{b}$ cross section
  ($R_\mathrm{b}$). Points with error bars are from
  Ref.~\cite{Aubert:2008ab}. The black solid curve is the result of
  the simultaneous fit to this distribution and the $\mbc$
  distributions (Fig.~\ref{mbc_fit_rel_ecm_240322}). The blue dashed
  curve is the visible cross section before accounting for the $\ecm$
  spread. The red dotted curve is the dressed cross section. The
  vertical dotted lines indicate the $B\bar{B}$ and $B\bar{B}^*$
  thresholds. }
\label{xsec_fit_rel_xsec_050422}
\end{figure}

The fit results are presented in Figs~\ref{mbc_fit_rel_ecm_240322} and
\ref{xsec_fit_rel_xsec_050422}, and in Table~\ref{tab:y4s_fit_results}.
\begin{table}[htbp]
\caption{ Results of the simultaneous fit to the $\mbc$ distributions
  in the $\Ufo$ data and the $R_b$ scan results of BaBar~\cite{Aubert:2008ab}. The lower
  three parameters are explained in the text. The first uncertainty is
  statistical, the second one (if present) is systematic.}
\renewcommand*{\arraystretch}{1.3}
\label{tab:y4s_fit_results}
\centering
\begin{tabular}{@{}ll@{}} \toprule
  Signal yield   & $(45.57\pm0.24\pm0.16)\times10^3$ \\
  Nominal $\ecm$ & $(10579.80\pm0.12\pm0.78)\,\mev$ \\  
  $\ecm$ spread  & $(5.62\pm0.19\pm0.27)\,\mev$ \\
  \midrule
  $\nr_3$ & $1.03^{+0.20}_{-0.23}$ \\
  $\sh_3$ & $(7.5^{+11.1}_{-\phantom{0}8.8})\,\mevc$ \\
  $\ff_3$ & $0.630^{+0.107}_{-0.101}$ \\
\bottomrule
\end{tabular}
\end{table}

\subsubsection{\boldmath Systematic uncertainties at the $\Ufo$}
\label{sec:y4s_syst}

We study systematic uncertainties from various sources
(Table~\ref{tab:y4s_syst}).
\begin{itemize}
\item We vary the $\bp$ and $\bn$ masses within their
  uncertainties. We use the Particle Data Group average values
  $m(B^+)=(5279.25\pm0.26)\,\mevm$ and
  $m(B^0)=(5279.63\pm0.20)\,\mevm$~\cite{ParticleDataGroup:2022pth}
  instead of their fit values because the latter include the
  measurement of $m(B^0)-m(B^+)$ at the $\Ufo$, performed by
  BaBar~\cite{BaBar:2008ikz}, that overlooks a source of systematic
  bias according to Ref.~\cite{Bondar:2022kxv}.
\item We use the cross section shape measured by
  Belle~\cite{Belle:2021lzm}. We find that $-2\ln{}\lik$ increases by
  11.0, which, for a model with 11 fewer parameters, corresponds to an
  exclusion level of $0.8\,\sigma$. Thus, the two shapes agree well.
\item We vary the $\deb$ shift by one standard deviation from
  Ref.~\cite{Belle:2021lzm}.
\item In the default fit, the corrections are applied only to
  the broken signal component in the sideband. We apply the
  corrections also to the broad broken signal component in
  the signal region assuming them to be the same.
\item The uncertainty due to the scale factor of
  Eq.~(\ref{eq:depr_fufa}) is found to be negligibly small.
\end{itemize}
In each case, the largest variation is taken as the uncertainty.  The
total systematic uncertainty is obtained by summing the individual
contributions in quadrature.
\begin{table}[htbp]
\caption{ Systematic uncertainty in the $B$-meson yield at the $\Ufo$,
  nominal $\ecm$ (in $\mev$), and $\ecm$ spread (in $\mev$). }
\renewcommand*{\arraystretch}{1.2}
\label{tab:y4s_syst}
\centering
\begin{tabular}{@{}lccc@{}} \toprule
  & \;\; $N$, $10^3$ \;\; & \;\; $\ecm$ \;\; & \;\; Spread \\
  \midrule
  $B^+$ mass                          & 0.01 & 0.47 & 0.04 \\
  $B^0$ mass                          & 0.01 & 0.32 & 0.09 \\
  Cross section shape                 & 0.05 & 0.10 & 0.02 \\
  Energy scale of the BaBar scan data & 0.03 & 0.53 & 0.25 \\
  Treatment of broken signal          & 0.15 & 0.02 & 0.00 \\
  \midrule
  Total                               & 0.16 & 0.78 & 0.27 \\
  \bottomrule
\end{tabular}
\end{table}

\subsection{\boldmath Determination of the FEI efficiency at the $\Ufo$}
\label{sec:fei_eff_y4s}

We determine the absolute value of the FEI efficiency at the $\Ufo$
using \eqref{eq:abs_eff_eq} to be
\begin{equation}
  \varepsilon = (0.5892 \pm 0.0031 \pm 0.0116) \times 10^{-3},
  \label{eq:abs_eff}
\end{equation}
where the first uncertainty is statistical and the second is systematic. 
The efficiency is higher than that at Belle,
$0.469\times10^{-3}$~\cite{Belle:2021lzm},
possibly due to Belle~II's higher reconstruction efficiency for
low-momentum charged particles.
The efficiency values at various energies determined from simulation
(Eq.~\eqref{eq:eff_vs_ecm}) are multiplied by a correction factor
$0.959\pm0.020$ which is the ratio of the value in
Eq.~\eqref{eq:abs_eff} and the constant term in
Eq.~\eqref{eq:eff_vs_ecm}. 

We consider contributions of the systematic uncertainties in $N$
(0.3\%) and $N_{\bb}$ (1.5\%). To check the stability of the
efficiency over the running period, we subdivide the $\Ufo$ sample
into three parts (one before the energy scan and two after) and
determine the efficiency in each. Based on the dispersion of these
measurements, we assign an additional 1.2\% systematic uncertainty due
to long-term variations. The total relative uncertainty in the FEI
efficiency, obtained by adding the statistical and systematic
contributions in quadrature (Eq.~(\ref{eq:abs_eff})), is 2.0\%.

\section{Measurements at the scan energies}
\label{sec:scan}

To determine the signal yields at various energies, we perform the
$\mbc$ fits as described in Section~\ref{sec:mbc_fits}. The energy
dependence of the $\ee\to\bb$, $\bbst$, and $\bstbst$ cross sections
influences the corresponding peak positions (due to the finite energy
spread~\cite{Bondar:2022kxv}) and ISR tails. Thus, energy-dependence
information is needed to determine the signal function in the $\mbc$
fit. The fit to the energy dependence is described in
Section~\ref{sec:fit_xsec_shape}. To obtain self-consistent results,
we use an iterative procedure: (1) we fit the $\mbc$ spectra; (2)
based on the signal yields, we determine the cross sections; and (3)
we fit the energy dependence of the cross sections.
We use the cross-section shapes measured by Belle as a starting
point~\cite{Belle:2021lzm}. Two iterations are sufficient for
convergence.
We report on the study of systematic uncertainties in
Section~\ref{sec:syst_at_scan}.

\subsection{\boldmath $\mbc$ fits at scan energies}
\label{sec:mbc_fits}

To fit the $\mbc$ distributions at the scan energies, we also include
the $\ee\to\bbst$ and $\ee\to\bstbst$ components.
Decays $B^*\to{B}\gamma$ lead to additional smearing of the $B$
momentum, which is accounted for in the fit
function~\cite{Belle:2021lzm}. The resulting $\mbc$ distribution is
sensitive to the distribution in the $B^*$ helicity angle, which is
the angle between the $B$ momentum in the $B^*$ rest frame and the
$B^*$ boost direction. For the $\ee\to\bbst$ process, the
helicity-angle distribution is fixed by conservation laws:
$1+\cos^2\theta_\mathrm{h}$. For the $\ee\to\bstbst$ process, the
distribution depends on an unknown parameter $\ah$:
$1+\ah\cos^2\theta_\mathrm{h}$, with $-1\leq\ah\leq1$; the parameter
$\ah$ is left free in the fits.

The $\ecm$ value and the $\ecm$ spread are determined by the fit for
all scan data samples.
The resolution-function parameters and the broken-signal
corrections ($\nr_3$, $\sh_3$, and $\ff_3$) are taken to be the same as
in the $\Ufo$ data (Tables~\ref{tab:y4s_mom_res} and
\ref{tab:y4s_fit_results}).

The fit interval is from $5.2\,\gevm$ up to the kinematic boundary at
$\ecm/2$. 
The smooth background is described by a square-root function
multiplied by a Chebyshev polynomial; we use third order for the two
higher scan energies and second order for the two lower ones. (A
larger fit interval requires a higher polynomial order; we find that
increasing the polynomial orders further does not significantly
improve the $-2\ln{}\lik$ of the fit.)

\begin{figure}[htbp]
\centering
\includegraphics[width=0.49\linewidth]{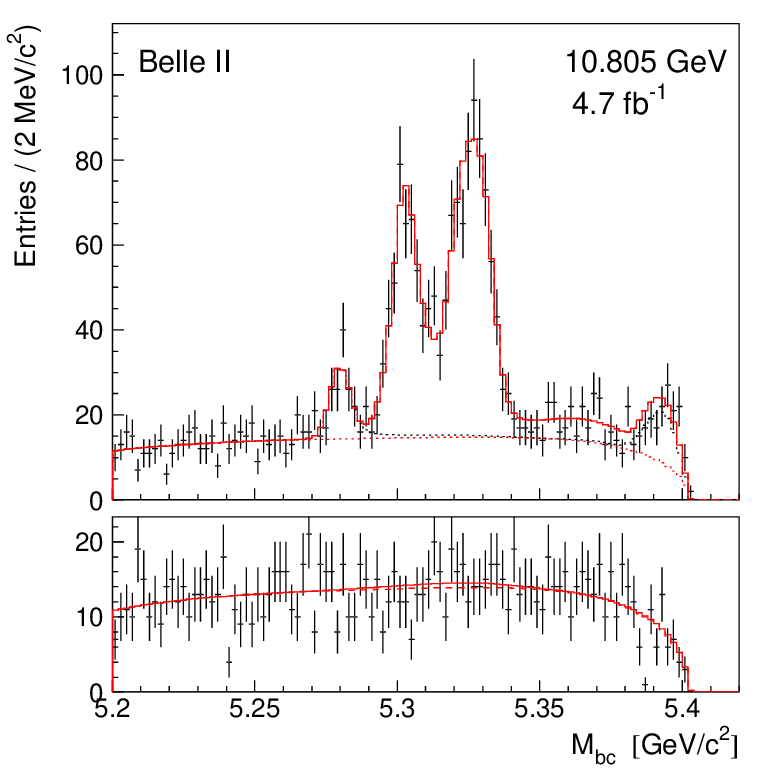}\hfill
\includegraphics[width=0.49\linewidth]{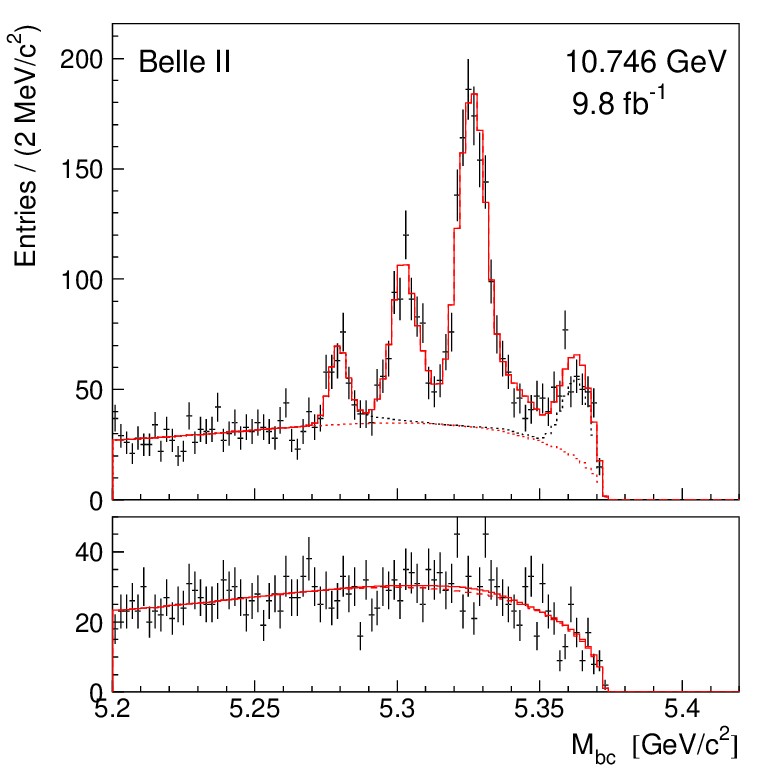}
\includegraphics[width=0.49\linewidth]{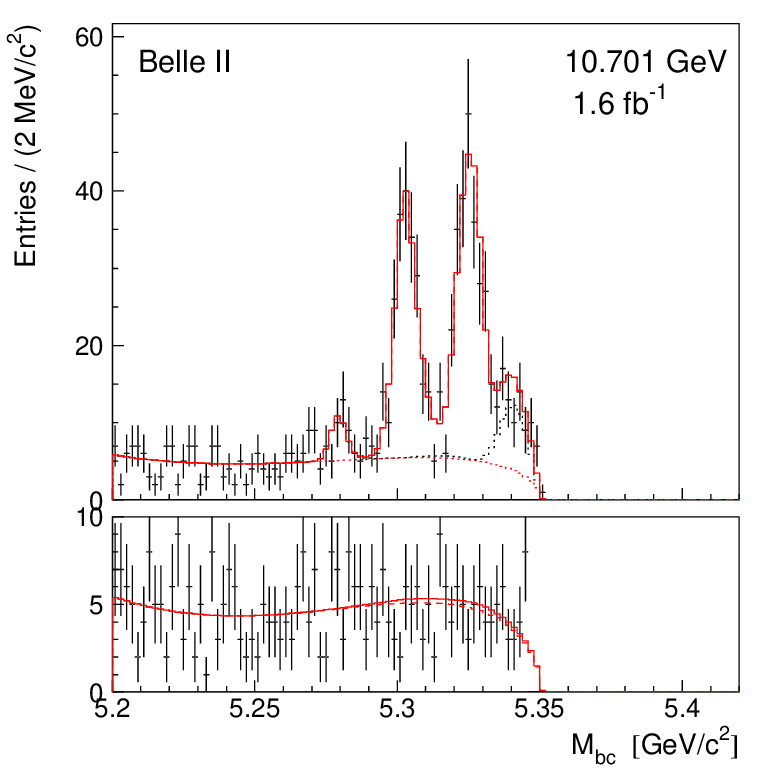}\hfill
\includegraphics[width=0.49\linewidth]{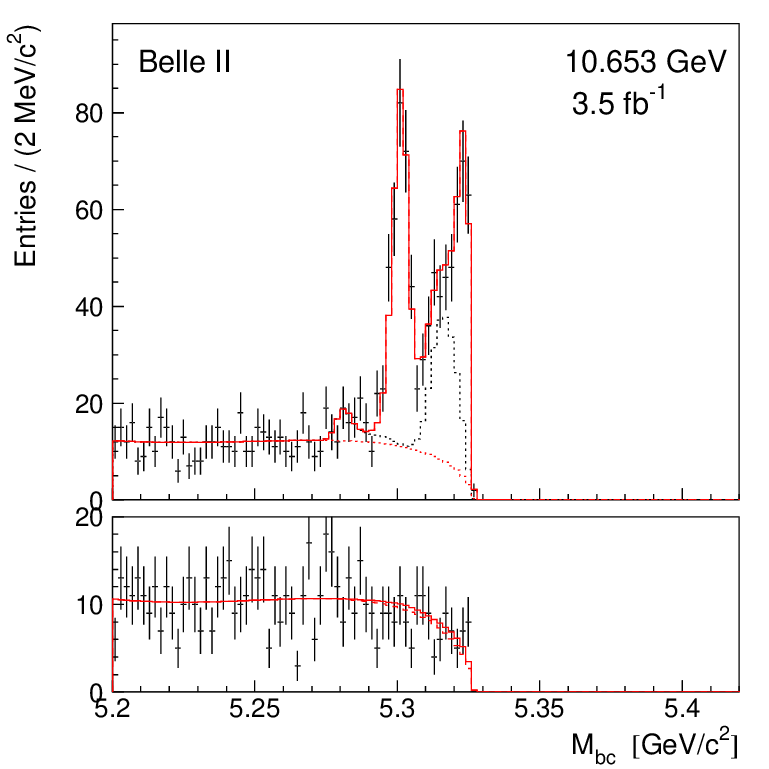}
\caption{ Distributions of $\mbc$ for the four scan energies. In each
  figure the top panel corresponds to the $\DE'$ signal region, and
  the bottom to the sideband. The red solid histogram shows the result
  of the fit; the red dashed histogram shows the smooth background;
  and the black dotted histogram shows the sum of the smooth
  background and the $\bb$ channel, which includes a peak near the
  threshold due to the ISR production of $\Ufo$. }
\label{mbc_scan_1_260123}
\end{figure}

Fit results for the scan data are shown in
Fig.~\ref{mbc_scan_1_260123}, and Tables~\ref{tab:lum_uncor_yi} and
\ref{tab:lum_ecm_spread}.
\begin{table}[htbp]
\caption{ Yields of the $\ee\to\bb$, $\bbst$, and $\bstbst$ processes
  in the scan data samples. The uncertainties are statistical. }
\renewcommand*{\arraystretch}{1.1}
\label{tab:lum_uncor_yi}
\centering
\begin{tabular}{@{}cccc@{}}
  \toprule
  point\# & $\bb$ & $\bbst$ & $\bstbst$ \\
  \midrule
  1 & $\phantom{1}90.1\pm17.5$ & $401.7\pm27.9$ & $525.4\pm30.8$ \\
  2 & $174.5\pm26.7$ & $535.6\pm42.6$ & $931.5\pm42.7$ \\
  3 & $\phantom{1}21.8\pm\phantom{1}8.2$ & $189.9\pm17.8$ & $202.5\pm17.9$ \\
  4 & $\phantom{1}32.2\pm14.6$ & $321.5\pm23.5$ & $151.4\pm15.0$ \\
  \bottomrule
\end{tabular}
\end{table}
\begin{table}[htbp]
\caption{ Results for the energy and energy spread of the scan data
  samples. The first uncertainty is statistical, the second is
  uncorrelated systematic, and the third is correlated systematic. }
\renewcommand*{\arraystretch}{1.1}
\label{tab:lum_ecm_spread}
\centering
\begin{tabular}{@{}ccc@{}}
  \toprule
  point\# & $\ecm\;(\mev)$ & $\ecm$ spread $(\mev)$ \\
  \midrule
  1 & $10804.50\pm0.65\pm0.25\pm0.50$ & $6.44\pm0.85\pm0.16\pm0.13$ \\
  2 & $10746.30\pm0.46\pm0.15\pm0.50$ & $5.68\pm0.69\pm0.29\pm0.09$ \\
  3 & $10700.90\pm0.61\pm0.14\pm0.50$ & $4.85\pm0.95\pm0.14\pm0.05$ \\
  4 & $10653.30\pm0.71\pm0.89\pm0.50$ & $5.23\pm0.57\pm0.66\pm0.20$ \\
  \bottomrule
\end{tabular}
\end{table}
The $\bb$ component has a peak near the kinematic boundary, which is
due to the ISR production of $\Ufo$ followed by decay to $\bb$. The
contribution of this process is fixed in the fits to the value
calculated using the energy dependence of the $\ee\to\bb$ dressed
cross-section between the $\bb$ and $\bbst$ thresholds, the
reconstruction efficiency, and the luminosity of each scan data
sample.
The data do not show any excess due to the three-body processes
$\ee\to{}B^{(*)}\bar{B}^{(*)}\pi$, whose signals are situated near the
$\mbc$ kinematic boundary and have shapes similar to that of the ISR
production of $\Ufo$. The absence of $B^{(*)}\bar{B}^{(*)}\pi$
signals is consistent with the observation that two-body processes
$B^{(*)}\bar{B}^{(*)}$ saturate the total $b\bar{b}$ cross section
below an energy of about $10.81\,\gev$ (see
Fig.~\ref{cmp_total_sum_exc_080323} below).
The yields are defined as the integrals of the signal components up to
$\mbc=5.35\,\gevm$.
At the two lowest scan energies, ISR production of the $\Ufo$ starts
to contribute to the above interval; the corresponding events are excluded
from the signal yield of $\ee\to\bb$.

\begin{figure}[htbp]
\centering
\includegraphics[width=0.32\linewidth]{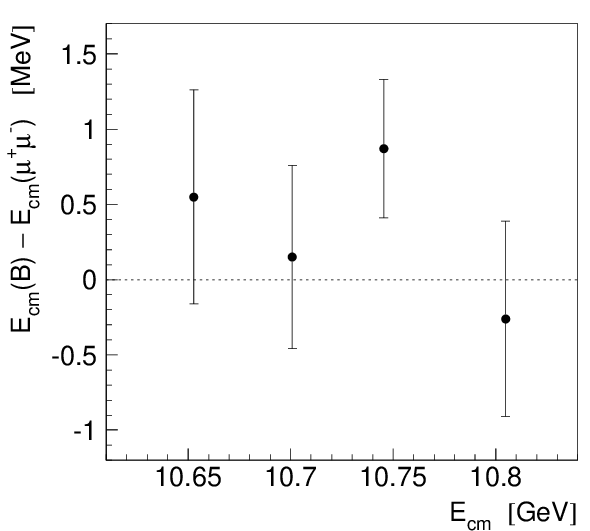}
\includegraphics[width=0.32\linewidth]{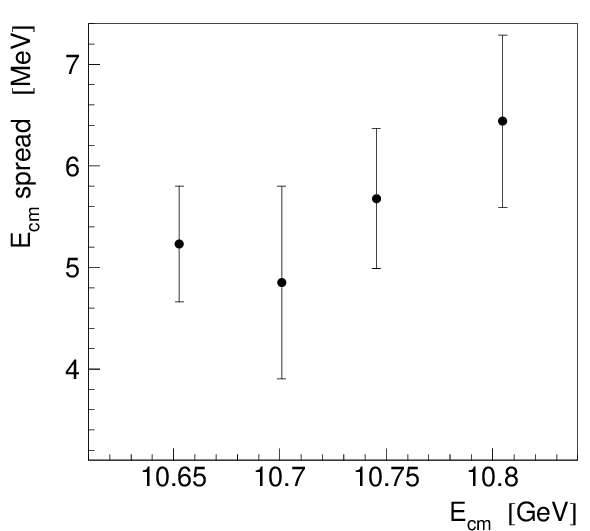}
\includegraphics[width=0.32\linewidth]{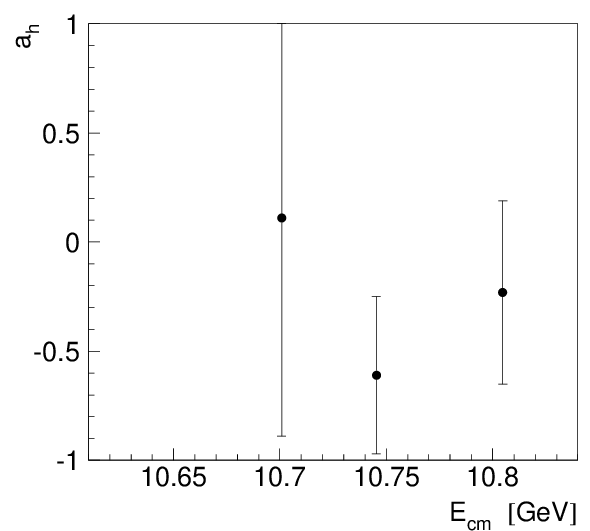}
\caption{ Measurements of $\ecm(B)-\ecm(\uu)$ (left), $\ecm$ spread
  (middle), and helicity parameter $\ah$ (right) at various
  energies. Error bars show statistical uncertainties. The
  uncorrelated and correlated uncertainties of $\ecm(\uu)$ are at the
  level of $0.3\,\mev$ and $1\,\mev$, respectively, and are not
  included. 
}
\label{decm_vs_ecm_260123}
\end{figure}

The measured $\ecm$ values agree with the results of the $\ee\to\uu$
analysis; the difference between the two measurements,
$\ecm(B)-\ecm(\uu)$, is shown in Fig.~\ref{decm_vs_ecm_260123}~(left).
The $\ecm$ spread values (Fig.~\ref{decm_vs_ecm_260123}~(right)) are
consistent across the scan data samples and agree with the $\Ufo$
measurement (Table~\ref{tab:y4s_fit_results}).
The results for the helicity parameter $\ah$
(Fig.~\ref{decm_vs_ecm_260123} (right)) are also consistent among the
scan data samples and agree with the value $-0.18\pm0.07$ measured by
Belle at the $\Uf$ energy~\cite{Belle:2021lzm}. As $\ecm$ decreases,
the $\bstbst$ signal width becomes smaller and sensitivity to $\ah$
drops. At the lowest $\ecm$ we set $\ah=0$; variations $\ah=-1$ and
$\ah=+1$ produce negligible changes in the yields.

We calculate the dressed cross sections as
\begin{equation}
  \sigma^\mathrm{dressed} = \frac{N}{\opdi\,L\,\varepsilon},
\end{equation}
where $N$ is the signal yield; $\opdi$ is the radiative correction,
calculated based on the cross-section shapes
(Fig.~\ref{xsec_vs_ecm_nice_260123} below) as described in
Section~\ref{sec:fit_y4s_data}; $L$ is the integrated luminosity; and
$\varepsilon$ is the reconstruction efficiency. The $\opdi$ values
are shown in Table~\ref{tab:lum_opdi}.
Results for the $\ee\to\bb$, $\ee\to\bbst$, and $\ee\to\bstbst$
processes are presented in Table~\ref{tab:lum_xsec} and
Fig.~\ref{xsec_vs_ecm_nice_260123}.
The cross sections match well with the previous measurement by
Belle~\cite{Belle:2021lzm} and have better precision.
\begin{table}[htbp]
\caption{ Values of $\opdi$ calculated using the cross-section shapes
  shown in Fig.~\ref{xsec_vs_ecm_nice_260123}.}
\renewcommand*{\arraystretch}{1.1}
\label{tab:lum_opdi}
\centering
\begin{tabular}{@{}cccc@{}}
  \toprule
  point\# & $\bb$ & $\bbst$ & $\bstbst$ \\
  \midrule
  1 & $0.757$ & $0.719$ & $0.646$ \\
  2 & $0.781$ & $0.860$ & $0.760$ \\
  3 & $0.696$ & $0.773$ & $0.686$ \\
  4 & $1.063$ & $0.720$ & $0.561$ \\
  \bottomrule
\end{tabular}
\end{table}
\begin{table}[htbp]
\caption{ Results for the $\ee\to\bb$, $\bbst$, and $\bstbst$ cross
  sections. The first uncertainty is statistical, the second is
  uncorrelated systematic and the third is correlated systematic.}
\renewcommand*{\arraystretch}{1.1}
\label{tab:lum_xsec}
\centering
\begin{tabular}{@{}cccc@{}}
  \toprule
  point\# & $\sigma(\ee\to\bb)$ (pb) & $\sigma(\ee\to\bbst)$ (pb) & $\sigma(\ee\to\bstbst)$ (pb) \\
  \midrule
  1 & $21.0\pm4.1\pm0.5\pm0.5$ & $\phantom{1}98.6\pm\phantom{1}6.8\pm1.3\pm2.3$ & $143.6\pm\phantom{1}8.4\pm2.4\pm3.3$ \\
  2 & $19.0\pm2.9\pm1.0\pm0.4$ & $\phantom{1}52.9\pm\phantom{1}4.2\pm1.0\pm1.2$ & $104.0\pm\phantom{1}4.8\pm2.7\pm2.4$ \\
  3 & $16.1\pm6.0\pm0.7\pm0.4$ & $126.0\pm11.8\pm2.2\pm2.8$ & $151.4\pm13.4\pm2.8\pm3.4$ \\
  4 & $\phantom{1}7.2\pm3.3\pm1.4\pm0.2$ & $106.8\pm\phantom{1}7.8\pm3.7\pm2.3$ & $\phantom{1}64.5\pm\phantom{1}6.4\pm2.7\pm1.4$ \\
  \bottomrule
\end{tabular}
\end{table}
\begin{figure}[htbp]
\centering
\includegraphics[width=0.49\linewidth]{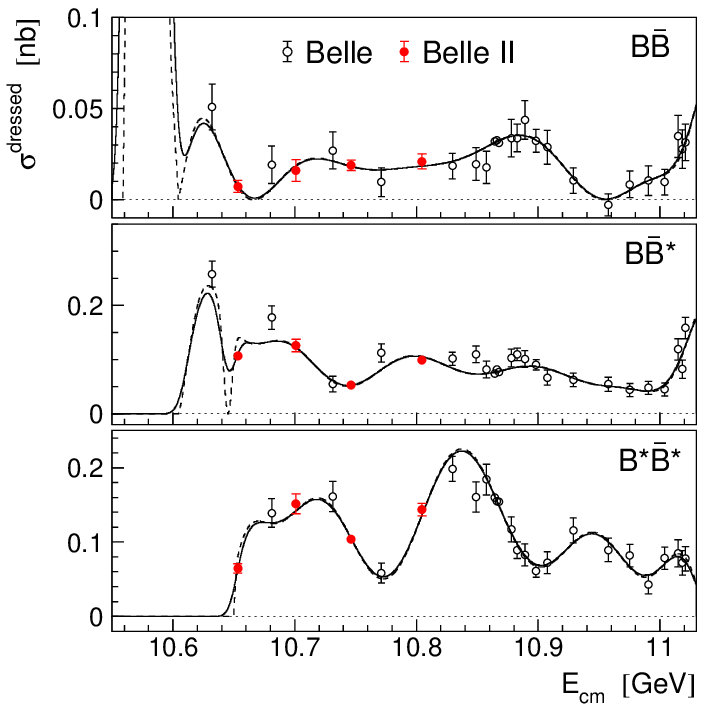}\hfill
\includegraphics[width=0.49\linewidth]{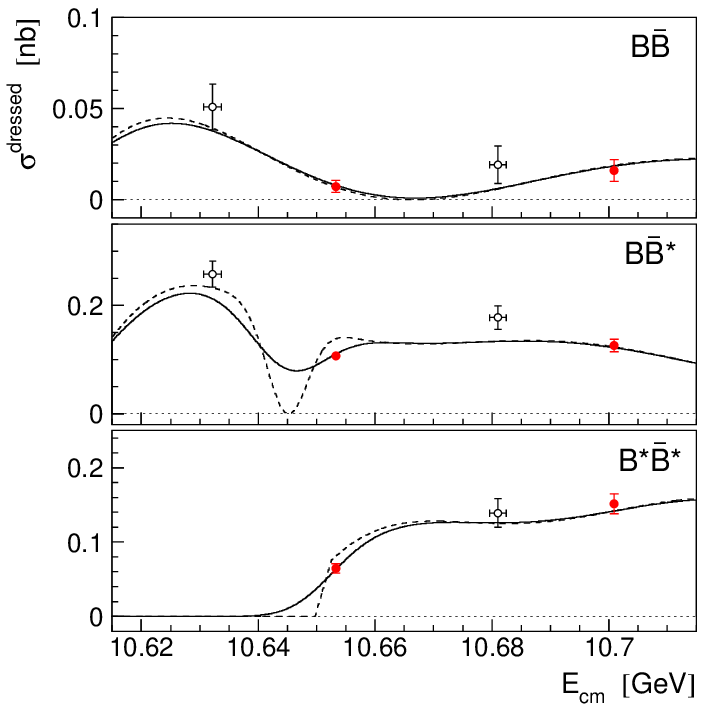}
\caption{ Energy dependence of the $\ee\to\bb$ (top), $\bbst$
  (middle), and $\bstbst$ (bottom) cross sections. Red circles show
  the Belle~II results obtained in this analysis; the error bars
  indicate the statistical uncertainty. The black open circles show
  the Belle measurements from Ref.~\cite{Belle:2021lzm}; the error
  bars show combined statistical and uncorrelated systematic
  uncertainties. The solid curves show the result of the fit to these
  three distributions and to the total $b\bar{b}$ cross section shown
  in Fig.~\ref{xtot_vs_ecm_260123}. The dashed curves show the fit
  function before the convolution accounting for the $\ecm$ spread.
  Plots on the left show the full fit range; on the right, the
  $\bstbst$ threshold region is magnified. }
\label{xsec_vs_ecm_nice_260123}
\end{figure}

\subsection{Fit to the energy dependence of the cross sections}
\label{sec:fit_xsec_shape}

\begin{figure}[htbp]
\centering
\includegraphics[width=0.53\linewidth]{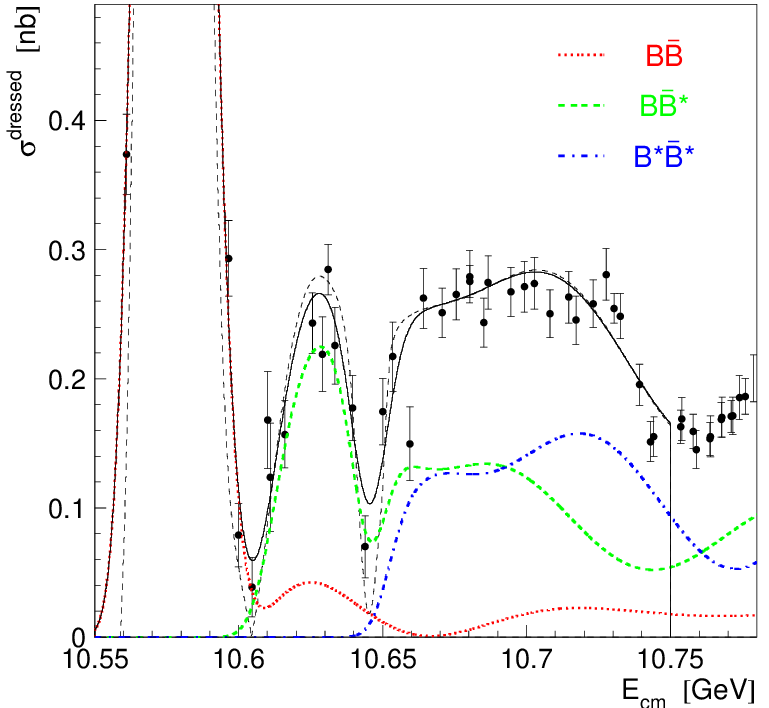}
\caption{ Energy dependence of the total $b\bar{b}$ dressed cross
  section from Ref.~\cite{Dong:2020tdw} (circles). The solid black curve
  shows the result of the simultaneous fit to this distribution and
  the exclusive $\bb$, $\bbst$, and $\bstbst$ cross section energy
  dependence (Fig.~\ref{xsec_vs_ecm_nice_260123}). The vertical line
  at $10.75\,\gev$ indicates the upper boundary of the fit
  interval. The dashed black curve shows the fit function before the
  convolution accounting for the $\ecm$ spread. Also shown are the
  individual contributions of $\bb$ (red dotted), $\bbst$ (green
  dashed), and $\bstbst$ (blue dash-dotted). }
\label{xtot_vs_ecm_260123}
\end{figure}

We perform a simultaneous fit to the energy dependence of the
exclusive $\ee\to\bb$, $\ee\to\bbst$, and $\ee\to\bstbst$ cross
sections, and the total $\ee\to{}b\bar{b}$ cross section. For the
exclusive cross sections, we use both the Belle~II results and the
Belle measurements from Ref.~\cite{Belle:2021lzm}. For the total cross
section, we use the data from Ref.~\cite{Dong:2020tdw}, where the
BaBar and Belle energy scan
results~\cite{Aubert:2008ab,Santel:2015qga} are combined and radiative
corrections are applied to convert the visible cross section into the
dressed one. The total cross section shows deep minima or zeros at the
$\bbst$ and $\bstbst$ thresholds
(Fig.~\ref{xtot_vs_ecm_260123}). These structures motivate inclusion
of the total cross section in the fit. The total cross section is
fitted only up to $10.75\,\gev$ (the $\bbst\pi$ threshold); its fit
function is the sum of the fit functions of the exclusive channels.
The $\Ufo$ peak region, where the dressed cross section reaches
approximately $2\,\mathrm{nb}$, is shown in more detail in
Fig.~\ref{cmp_total_sum_exc_080323} below.

To fit the $\ee\to\bb$ cross section we use a 13th order Chebyshev
polynomial.
Since there is a zero at the $\bbst$ threshold
($E_{\bbst}=10604.3\,\mev$) in the fit shown in
Fig.~\ref{xsec_fit_rel_xsec_050422}, the cross section is fitted only
above the $\bbst$ threshold.
We impose the requirement that the polynomial
is zero at the $\bbst$ threshold by adding a point there with zero
cross section and a small uncertainty. Below the $\bbst$ threshold, we
use the cross-section shape obtained in the analysis of the $\Ufo$
data (Fig.~\ref{xsec_fit_rel_xsec_050422}).

To fit the $\ee\to\bbst$ cross section we use a 14th order Chebyshev
polynomial plus a Gaussian
$A_0\exp\left\{-(x-E_0)^2/2\sigma_0^2\right\}$ to describe the dip at
the $\bstbst$ threshold ($E_{\bstbst}=10649.7\,\mev$). We impose a
requirement that the polynomial is equal to zero at the $\bbst$
threshold. The parameters of the Gaussian are free in the fit. We find
$A_0=(-0.180^{+0.039}_{-0.018})\,\mathrm{nb}$,
$E_0=(10644.9^{+1.2}_{-1.4})\,\mev$, and
$\sigma_0=(3.5^{+1.4}_{-0.5})\,\mev$.

The shape of the $\ee\to\bstbst$ cross section is parameterized using
a 13th order Chebyshev polynomial and a linear function at the
threshold. The data in Figs.~\ref{xsec_vs_ecm_nice_260123} and
\ref{xtot_vs_ecm_260123} show that the $\ee\to\bstbst$ cross section
increases rapidly above the $\bstbst$ threshold. To describe this
increase, we use the function
\begin{equation}
  f_1(\ecm) = \frac{1}{k}\,(\ecm-E_{\bstbst}).
\end{equation}
We do not impose a requirement of a zero for the high-order polynomial,
and the fit function is equal to whichever is smaller: $f_1(\ecm)$ or
the high-order polynomial.
The fit to the energy dependence of the cross sections is not
sensitive to the slope of $f_1(\ecm)$.
However, the shape of the $\mbc$ distribution at the energy
$\ecm=10653.3\,\mev$ (only about $4\,\mev$ above the $\bstbst$
threshold) is sensitive to the slope.
The parameter $k$ is free in the corresponding fit
(Fig.~\ref{mbc_scan_1_260123} (bottom right)) and equals
$k = (37 \pm 13)\,\mev / \mathrm{nb}$.
This shows that the $\bstbst$ cross section
reaches the typical level of $0.1\,\mathrm{nb}$ within $3.7\,\mev$
from the threshold. Thus, the shape of the $\mbc$ distribution
provides further support for the rapid rise of the $\bstbst$ cross
section.

The degrees of the polynomials chosen are the lowest that provide
reasonable descriptions of the shape, while also allowing for
variations in degree to estimate systematic uncertainties.

\subsection{Systematic uncertainties}
\label{sec:syst_at_scan}

\begin{figure}[htbp]
\centering
\includegraphics[width=0.44\linewidth]{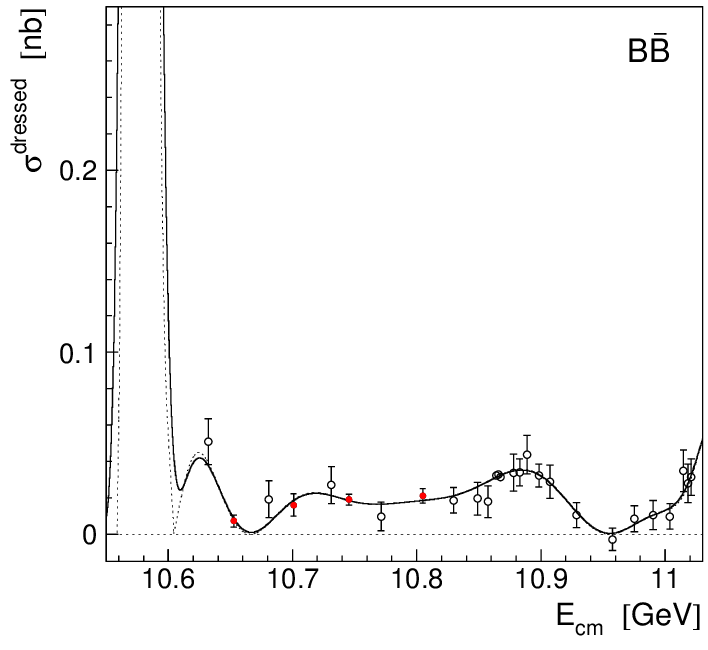}\hfill
\includegraphics[width=0.44\linewidth]{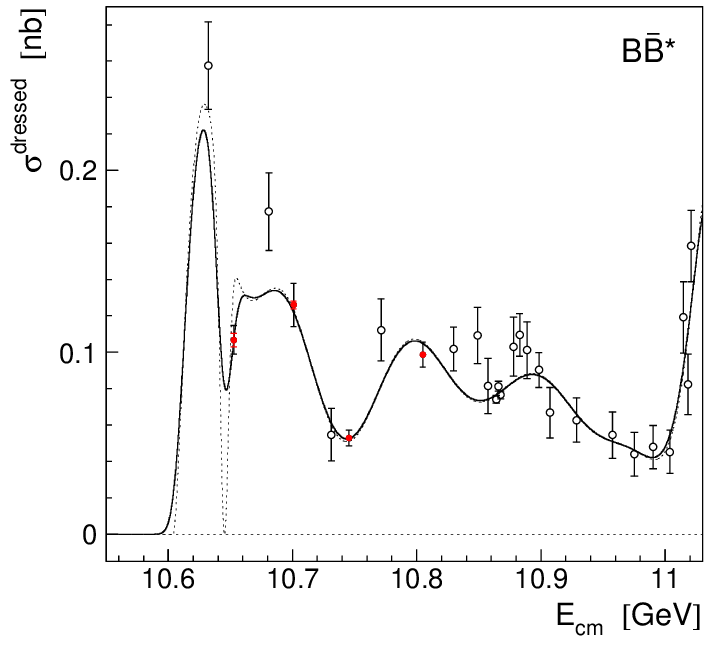}
\includegraphics[width=0.44\linewidth]{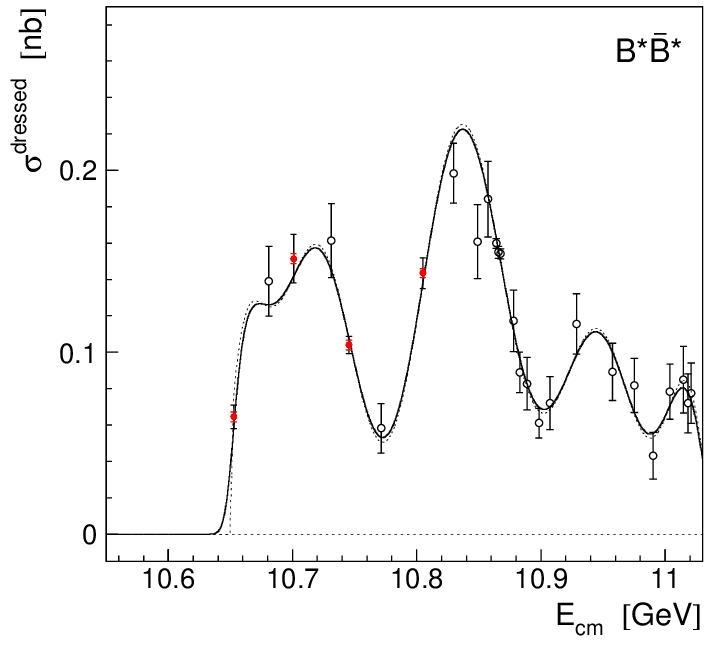}\\
\caption{ Energy dependence of the $\ee\to\bb$ (left), $\bbst$
  (right), and $\bstbst$ (bottom) cross sections. Red filled and black
  open circles show the Belle~II and Belle measurements,
  respectively. In the Belle~II results, black error bars show
  statistical uncertainties, red error bars show uncorrelated
  systematic uncertainties. (On six out of the twelve points, the
  systematic uncertainties are smaller than the size of the red filled
  circle, and thus not visible.) The solid curves show the default fit
  result; the dashed curves show the fit function before the
  convolution used to account for the $\ecm$ spread. }
\label{bb_xs_vs_ecm_syst_unc_260123}
\end{figure}

\begin{figure}[htbp]
\centering
\includegraphics[width=0.33\linewidth]{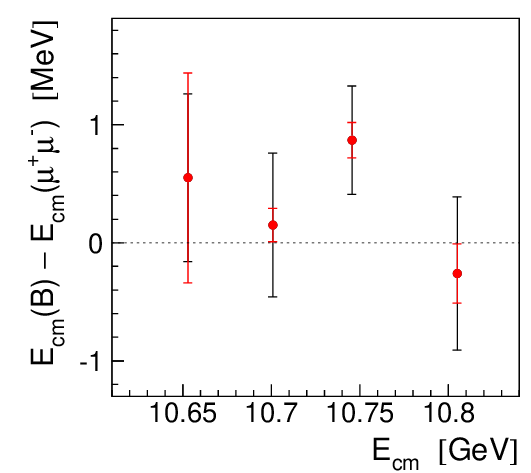}\hspace{1cm}
\includegraphics[width=0.33\linewidth]{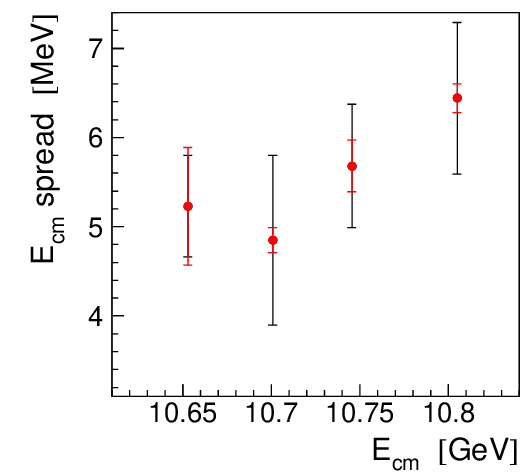}
\caption{ Measurements of the energy difference $\ecm(B)-\ecm(\uu)$
  (left) and the $\ecm$ spread (right) at various energies. Black
  error bars show statistical uncertainties, red error bars show
  uncorrelated systematic uncertainties. }
\label{decm_vs_ecm_syst_unc_130323}
\end{figure}

We consider the following sources of uncorrelated systematic
uncertainty:
\begin{itemize}
\item Cross section shape --- model: we change the orders of the
  Chebyshev polynomials used to parameterize the cross section shapes
  by $\pm1$, $+2$, and $+3$ in all three channels simultaneously. We
  repeat the $\mbc$ fits using the new cross-section shapes and take
  the root-mean-square (RMS) deviation of the measured quantities as
  the uncertainty. These are small compared to the statistical
  uncertainties.
\item Cross section shape --- uncertainty in measurements: we use
  pseudoexperiments generated using the fitted values of the cross
  sections and the uncertainties found in data. The Belle measurements
  of the exclusive cross sections and the measurements of the total
  cross section are also varied. For each pseudoexperiment, we fit the
  energy dependence of the cross sections and perform the $\mbc$ fits
  using the resulting cross-section shapes. The RMS deviation of the
  measured quantities is taken as the uncertainty. This source is
  found to give the largest contribution; however, it is small
  compared to the statistical uncertainties.
  Figure~\ref{bb_xs_vs_ecm_syst_xsec_stat_260123} shows examples of
  fits to the pseudoexperiments and the systematic uncertainties from
  this source.
\begin{figure}[htbp]
\centering
\includegraphics[width=0.49\linewidth]{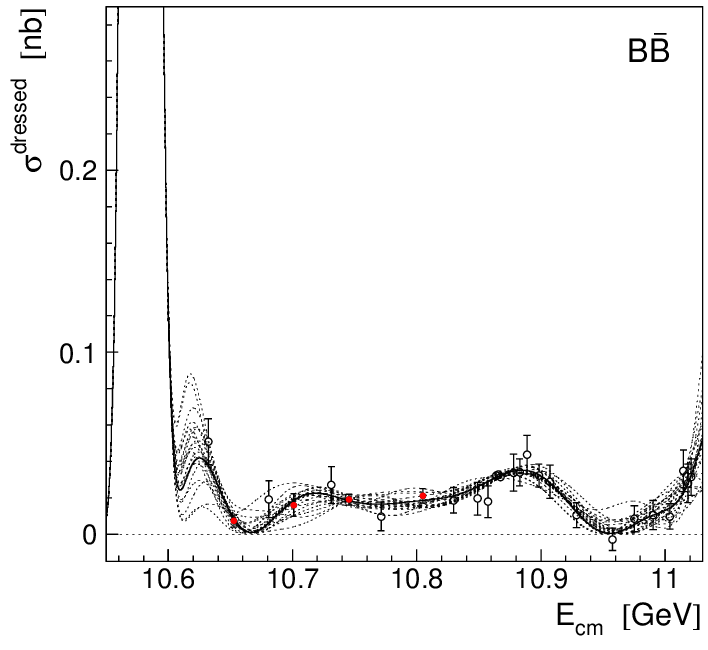}\hfill
\includegraphics[width=0.49\linewidth]{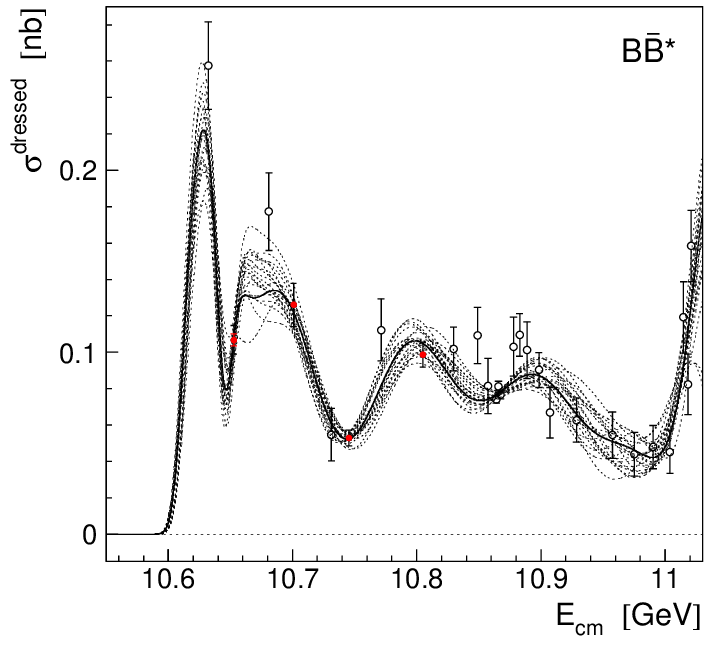}
\includegraphics[width=0.49\linewidth]{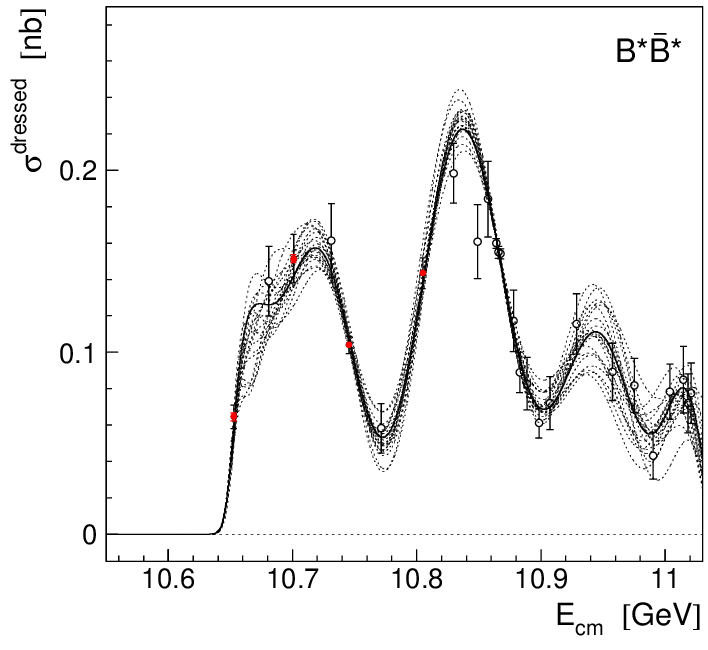}\\
\caption{ Energy dependence of the $\ee\to\bb$ (left), $\bbst$
  (right), and $\bstbst$ (bottom) cross sections. Red filled and black
  open circles show the Belle~II and Belle measurements,
  respectively. In the Belle~II results, black error bars show
  statistical uncertainties and red error bars show systematic
  uncertainties related to the shape of the cross section energy
  dependence, estimated using pseudoexperiments. (On nine out of the
  twelve points, the systematic uncertainties are smaller than the
  size of the red filled circle, and thus not visible.) The solid
  curves show the default fit result; the dashed curves show the
  results of the fits to various pseudoexperiments. }
\label{bb_xs_vs_ecm_syst_xsec_stat_260123}
\end{figure}
\item Broken signal: we consider the same variation of the fit as
  described above for the $\Ufo$ analysis
  (Section~\ref{sec:y4s_syst}). Namely, in the default fit the
  corrections are applied only to the broken signal component in the
  sideband. We apply the corrections also to the broad broken signal
  component in the signal region assuming that they are the same as
  for the broken signal component in the sideband. The uncertainties
  due to this source are small.
\item Shape of the smooth background: in the default fit the smooth
  background is described by a threshold function multiplied by 2nd or
  3rd order Chebyshev polynomial (Section~\ref{sec:mbc_fits}). We
  increase the order of the polynomial by one and two units. The
  uncertainties due to this source are small.
\end{itemize}
The total uncorrelated systematic uncertainty is obtained by adding
the above contributions in quadrature; it is shown in
Figs.~\ref{bb_xs_vs_ecm_syst_unc_260123} and
\ref{decm_vs_ecm_syst_unc_130323}, and in
Tables~\ref{tab:lum_ecm_spread} and \ref{tab:lum_xsec}.
For simplicity, in the default fit to the cross section energy
dependence, we use only statistical uncertainties of the Belle~II
measurements. We repeat the fit combining statistical and uncorrelated
systematic uncertainties for Belle~II and find that the change in the
results is negligibly small.

Correlated systematic uncertainty in the cross sections has the
following sources:
\begin{itemize}
\item Uncertainty in the absolute value of the efficiency of 2.1\%
  determined using the $\Ufo$ data (Section~\ref{sec:fei_eff_y4s}).
\item Uncertainty in the energy dependence of the efficiency. For the
  efficiency, we use the values of the fit function shown in
  Fig.~\ref{eff_vs_ecm_280322}. The uncertainty due to the limited
  size of the simulated samples has values in the range $0.3\%-0.8\%$,
  increasing linearly with $\ecm$.
\item Uncertainty in the luminosity of 0.6\%~\cite{Belle-II:2019usr}.
\end{itemize}
The effect of the uncertainties in the $B$ and $B^*$ masses on the
cross sections is negligibly small.
The total correlated systematic uncertainties shown in
Table~\ref{tab:lum_xsec} are obtained by adding the above contributions
in quadrature.

The uncertainties in the $B$ and $B^*$ masses lead to correlated
uncertainties in $\ecm$ and in the $\ecm$ spread. The corresponding
contributions are determined by varying the masses by their
uncertainties and repeating the analysis. The uncertainties in $\ecm$
are shown in Table~\ref{tab:syst_ecm_masses}. They are roughly
independent of the energy. The total uncertainty is calculated as the
sum in quadrature of the various contributions.
The uncertainties in the $\ecm$ spread are relatively small; their
total contributions are shown in Table~\ref{tab:lum_ecm_spread}.
\begin{table}[htbp]
\caption{ Systematic uncertainty in $\ecm$ due to uncertainties in the
  $B$ and $B^*$ masses. }
\renewcommand*{\arraystretch}{1.1}
\label{tab:syst_ecm_masses}
\centering
\begin{tabular}{@{}lc@{}}
  \toprule
  Source & Systematic uncertainty $(\mev)$ \\
  \midrule
  $m(B^+) = (5279.25 \pm 0.26)\,\mevm$      & $\pm0.29$ \\
  $m(B^0) = (5279.63 \pm 0.20)\,\mevm$      & $\pm0.18$ \\
  $m(B^*) - m(B) = (45.42 \pm 0.26)\,\mevm$ & $\pm0.37$ \\
  \midrule
  Total                                     & $\pm0.50$ \\
  \bottomrule
\end{tabular}
\end{table}

\section{Discussion and summary}
\label{sec:discussion}

\begin{figure}[htbp]
  \centering
  \includegraphics[width=0.49\textwidth]{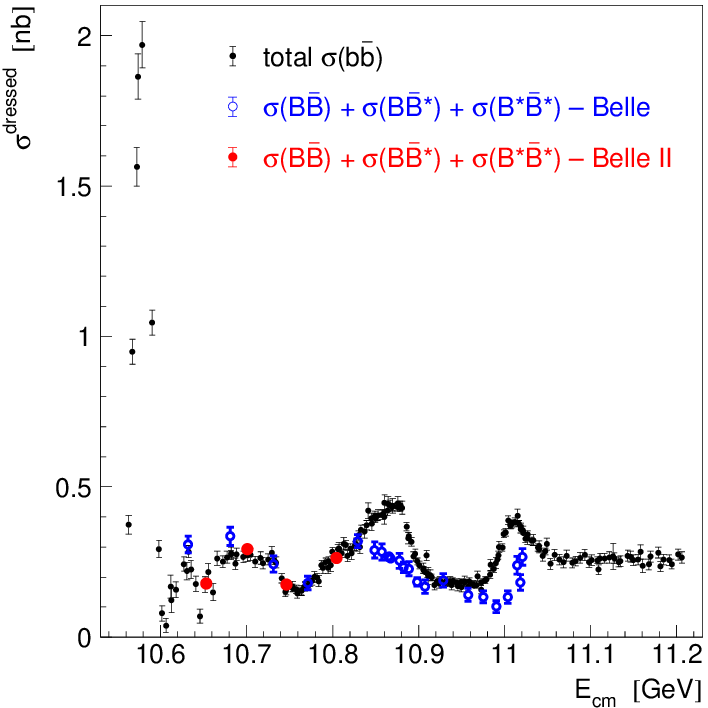}\hfill
  \includegraphics[width=0.49\textwidth]{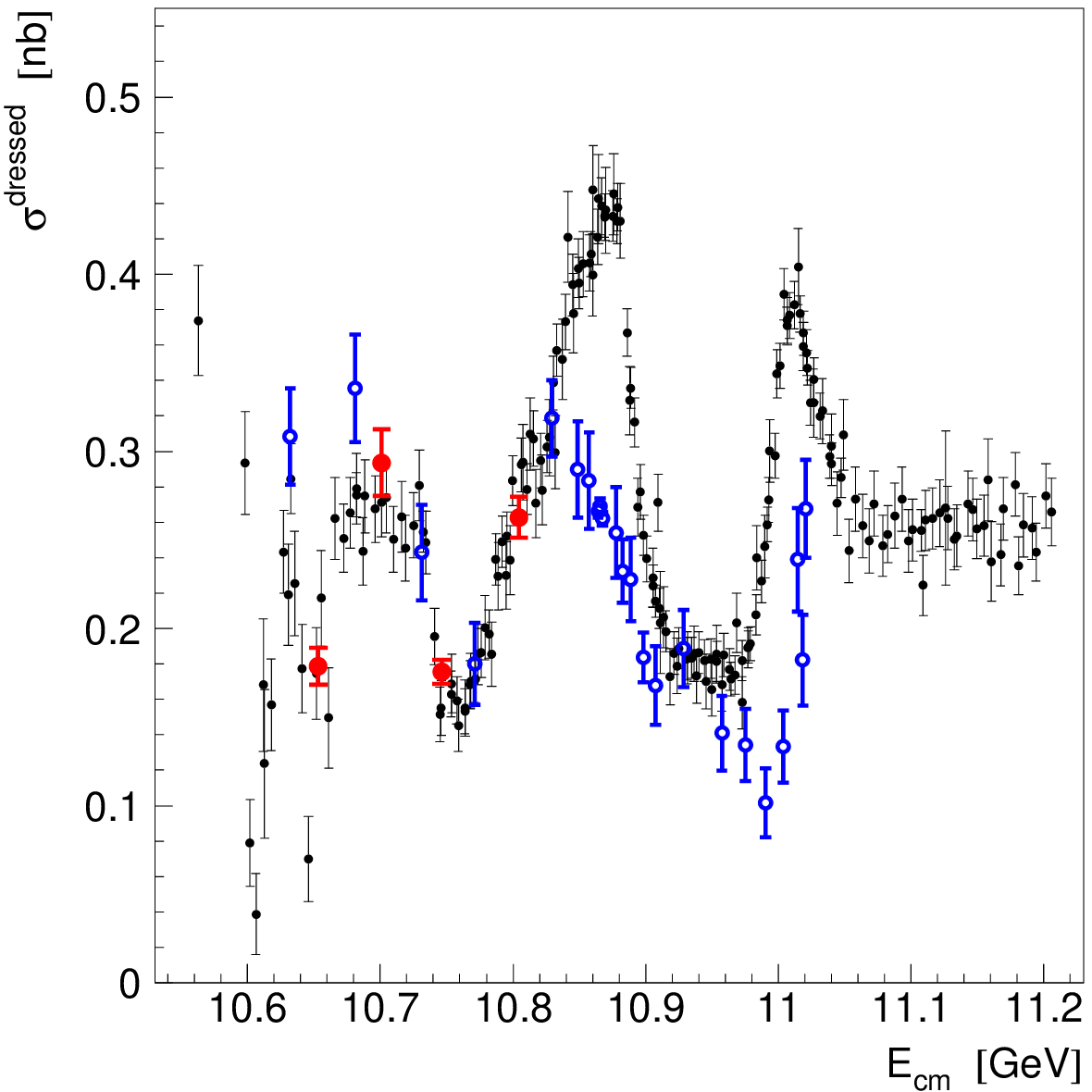}
  \caption{ Energy dependence of the total $b\bar{b}$ dressed cross
    section obtained in Ref.~\cite{Dong:2020tdw} from the visible
    cross sections measured by Belle~\cite{Santel:2015qga}, and
    BaBar~\cite{Aubert:2008ab} (black circles) and the sum of the
    exclusive $\bb$, $\bbst$, and $\bstbst$ cross sections measured by
    Belle~\cite{Belle:2021lzm} (open blue circles) and in this work
    (filled red circles). The right panel shows the low cross-section
    region with an expanded scale. }
  \label{cmp_total_sum_exc_080323}
\end{figure}

In Fig.~\ref{cmp_total_sum_exc_080323} we show the sum of the
exclusive $\bb$, $\bbst$, and $\bstbst$ cross sections measured in
this work and in the Belle experiment~\cite{Belle:2021lzm},
superimposed on the total $b\bar{b}$ dressed cross
section~\cite{Dong:2020tdw}.
The sum of measurements performed in this work agrees well with the
total cross section up to $\ecm = 10.84\,\gev$. 
The deviation at higher energy is presumably due to the contribution
of $\bs$ mesons, multibody final states
$B^{(*)}\bar{B}^{(*)}\pi(\pi)$, and production of bottomonia in
association with light hadrons.

\begin{figure}[htbp]
\centering
\includegraphics[width=0.49\linewidth]{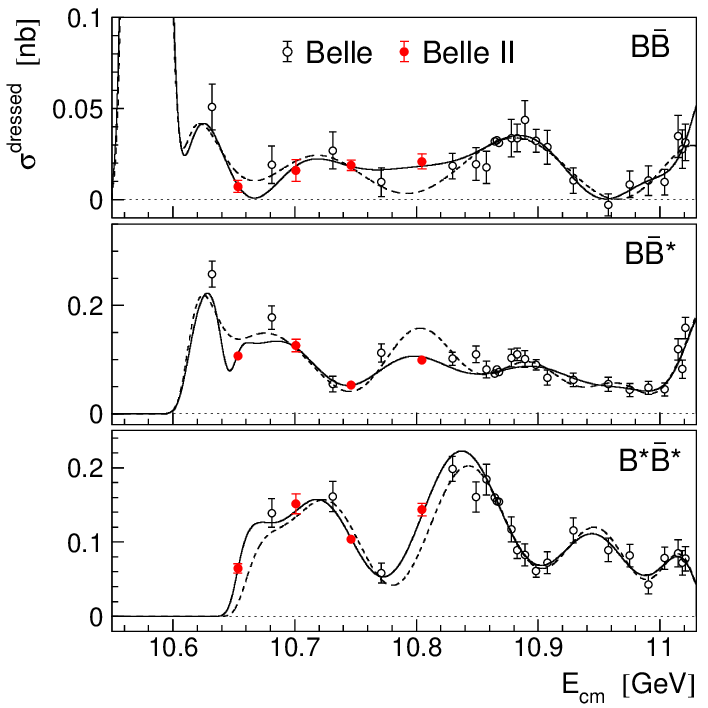}
\caption{ Energy dependence of the cross sections: $\ee\to\bb$ (top),
  $\bbst$ (middle) and $\bstbst$ (bottom). Symbols and curves have the
  same meaning as in Fig.~\ref{xsec_vs_ecm_nice_260123}, except that
  dashed curves show the result of the fit to the Belle points
  only~\cite{Belle:2021lzm}. }
\label{xsec_vs_ecm_ini_061022}
\end{figure}

The results of the fit to the Belle and Belle~II points, and a
separate fit to the Belle points only, are shown in
Fig.~\ref{xsec_vs_ecm_ini_061022}.
As measurements of the dressed cross section rely on the energy
dependence of the cross section as an input, and our measurements have
improved the knowledge of this energy dependence, the Belle
measurements may be shifted as a result. However, we find that the
shifts are negligibly small compared to the corresponding
uncertainties.

A new and unexpected observation of this analysis is that the
$\ee\to\bstbst$ cross section increases very rapidly just above the
threshold. The energy of the nearby
scan point $\ecm=(10653.30\pm1.14)\,\mev$ is only
$(2.96\pm1.52)\,\mev$ higher than the $B^{*0}\bar{B}^{*0}$ threshold
and only $(4.78\pm1.47)\,\mev$ higher than the $B^{*+}B^{*-}$
threshold; these differences are less than the $\ecm$ spread of 
$\sigma=(5.23\pm0.89)\,\mev$.
The large observed value of the $\bstbst$ cross section at this scan
point is especially surprising since phase space in the reaction
$\ee\to\bstbst$ grows as the $3/2$ power of the difference between the
beam energy and the threshold energy, and thus the derivative of the
cross section of this process must vanish at the threshold. Hence, we
conclude that the amplitude of the process under consideration
increases rapidly towards the threshold.
This phenomenon can be explained by the presence of a $\bstbst$
molecular state near the $\bstbst$ threshold.
Such a state could be bound or virtual; the $B^*$ and $\bar{B}^*$
mesons are in relative P-wave. There is a similar phenomenon near
$D^*\bar{D}^*$ threshold, which was explained in
Ref.~\cite{Dubynskiy:2006sg} by the presence of a $D^*\bar{D}^*$ molecule.

The existence of a $\bstbst$ molecule would also provide a natural
explanation for the narrow dip in the $\ee\to\bbst$ cross section near
the $\bstbst$ threshold (Fig.~\ref{xsec_vs_ecm_nice_260123}), as
destructive interference between $\ee\to\bbst$ and the
$\ee\to\bstbst\to\bbst$ rescattering process.\footnote{Alternatively,
  the dip may arise from a node in the $\Upsilon(4S)$ wave
  function~\cite{Ono:1985eu}.}
One can expect that such a complex behavior of the cross sections near
the $\bstbst$ threshold might lead to nontrivial effects, such as a
large violation of isospin symmetry, enhancement of inelastic
processes such as $\ee\to\U(nS)\pp$ and $h_b(1P)\eta$, and violation
of heavy-quark spin symmetry due to the interaction of $B$ mesons in
the final state~\cite{Li:2013kya}.\footnote{Phenomenological
  analysis~\cite{Salnikov:2023cug} of the preliminary
  results~\cite{Bertemes:2023LaThuile} of this work confirms the
  presence of a molecular state near the $\bstbst$ threshold.} Further
study of this energy region with larger samples should make it
possible to confirm or refute the existence of such a near-threshold
resonance.

In summary, we report measurements of the $\ee\to\bb$, $\ee\to\bbst$,
and $\ee\to\bstbst$ cross sections at four energies between $10.65$
and $10.80\,\gev$ using Belle~II data with a total integrated
luminosity of $19.7\,\fb$ (Table~\ref{tab:lum_xsec}).  The obtained
two-body cross sections at these energies are consistent with the
earlier results of Belle~\cite{Belle:2021lzm}, and provide significant
additional information. These results can be used in coupled-channel
analysis of energy-scan data to extract parameters of the highly
excited $\Upsilon$ states, in particular, of the recently observed
$\U(10753)$ state. We find that the $\ee\to\bstbst$ cross section
increases very rapidly above the corresponding threshold, which might
indicate the presence of a $\bstbst$ molecular state near the
threshold.

\section{Acknowledgements}

This work, based on data collected using the Belle II detector, which was built and commissioned prior to March 2019, was supported by
Higher Education and Science Committee of the Republic of Armenia Grant No.~23LCG-1C011;
Australian Research Council and Research Grants
No.~DP200101792, 
No.~DP210101900, 
No.~DP210102831, 
No.~DE220100462, 
No.~LE210100098, 
and
No.~LE230100085; 
Austrian Federal Ministry of Education, Science and Research,
Austrian Science Fund
No.~P~34529,
No.~J~4731,
No.~J~4625,
and
No.~M~3153,
and
Horizon 2020 ERC Starting Grant No.~947006 ``InterLeptons'';
Natural Sciences and Engineering Research Council of Canada, Compute Canada and CANARIE;
National Key R\&D Program of China under Contract No.~2022YFA1601903,
National Natural Science Foundation of China and Research Grants
No.~11575017,
No.~11761141009,
No.~11705209,
No.~11975076,
No.~12135005,
No.~12150004,
No.~12161141008,
and
No.~12175041,
and Shandong Provincial Natural Science Foundation Project~ZR2022JQ02;
the Czech Science Foundation Grant No.~22-18469S 
and
Charles University Grant Agency project No.~246122;
European Research Council, Seventh Framework PIEF-GA-2013-622527,
Horizon 2020 ERC-Advanced Grants No.~267104 and No.~884719,
Horizon 2020 ERC-Consolidator Grant No.~819127,
Horizon 2020 Marie Sklodowska-Curie Grant Agreement No.~700525 ``NIOBE''
and
No.~101026516,
and
Horizon 2020 Marie Sklodowska-Curie RISE project JENNIFER2 Grant Agreement No.~822070 (European grants);
L'Institut National de Physique Nucl\'{e}aire et de Physique des Particules (IN2P3) du CNRS
and
L'Agence Nationale de la Recherche (ANR) under grant ANR-21-CE31-0009 (France);
BMBF, DFG, HGF, MPG, and AvH Foundation (Germany);
Department of Atomic Energy under Project Identification No.~RTI 4002,
Department of Science and Technology,
and
UPES SEED funding programs
No.~UPES/R\&D-SEED-INFRA/17052023/01 and
No.~UPES/R\&D-SOE/20062022/06 (India);
Israel Science Foundation Grant No.~2476/17,
U.S.-Israel Binational Science Foundation Grant No.~2016113, and
Israel Ministry of Science Grant No.~3-16543;
Istituto Nazionale di Fisica Nucleare and the Research Grants BELLE2;
Japan Society for the Promotion of Science, Grant-in-Aid for Scientific Research Grants
No.~16H03968,
No.~16H03993,
No.~16H06492,
No.~16K05323,
No.~17H01133,
No.~17H05405,
No.~18K03621,
No.~18H03710,
No.~18H05226,
No.~19H00682, 
No.~20H05850,
No.~20H05858,
No.~22H00144,
No.~22K14056,
No.~22K21347,
No.~23H05433,
No.~26220706,
and
No.~26400255,
and
the Ministry of Education, Culture, Sports, Science, and Technology (MEXT) of Japan;  
National Research Foundation (NRF) of Korea Grants
No.~2016R1\-D1A1B\-02012900,
No.~2018R1\-A2B\-3003643,
No.~2018R1\-A6A1A\-06024970,
No.~2019R1\-I1A3A\-01058933,
No.~2021R1\-A6A1A\-03043957,
No.~2021R1\-F1A\-1060423,
No.~2021R1\-F1A\-1064008,
No.~2022R1\-A2C\-1003993,
and
No.~RS-2022-00197659,
Radiation Science Research Institute,
Foreign Large-Size Research Facility Application Supporting project,
the Global Science Experimental Data Hub Center of the Korea Institute of Science and Technology Information
and
KREONET/GLORIAD;
Universiti Malaya RU grant, Akademi Sains Malaysia, and Ministry of Education Malaysia;
Frontiers of Science Program Contracts
No.~FOINS-296,
No.~CB-221329,
No.~CB-236394,
No.~CB-254409,
and
No.~CB-180023, and SEP-CINVESTAV Research Grant No.~237 (Mexico);
the Polish Ministry of Science and Higher Education and the National Science Center;
the Ministry of Science and Higher Education of the Russian Federation
and
the HSE University Basic Research Program, Moscow;
University of Tabuk Research Grants
No.~S-0256-1438 and No.~S-0280-1439 (Saudi Arabia);
Slovenian Research Agency and Research Grants
No.~J1-9124
and
No.~P1-0135;
Agencia Estatal de Investigacion, Spain
Grant No.~RYC2020-029875-I
and
Generalitat Valenciana, Spain
Grant No.~CIDEGENT/2018/020;
The Knut and Alice Wallenberg Foundation (Sweden), Contracts No.~2021.0174 and No.~2021.0299;
National Science and Technology Council,
and
Ministry of Education (Taiwan);
Thailand Center of Excellence in Physics;
TUBITAK ULAKBIM (Turkey);
National Research Foundation of Ukraine, Project No.~2020.02/0257,
and
Ministry of Education and Science of Ukraine;
the U.S. National Science Foundation and Research Grants
No.~PHY-1913789 
and
No.~PHY-2111604, 
and the U.S. Department of Energy and Research Awards
No.~DE-AC06-76RLO1830, 
No.~DE-SC0007983, 
No.~DE-SC0009824, 
No.~DE-SC0009973, 
No.~DE-SC0010007, 
No.~DE-SC0010073, 
No.~DE-SC0010118, 
No.~DE-SC0010504, 
No.~DE-SC0011784, 
No.~DE-SC0012704, 
No.~DE-SC0019230, 
No.~DE-SC0021274, 
No.~DE-SC0021616, 
No.~DE-SC0022350, 
No.~DE-SC0023470; 
and
the Vietnam Academy of Science and Technology (VAST) under Grants
No.~NVCC.05.12/22-23
and
No.~DL0000.02/24-25.

These acknowledgements are not to be interpreted as an endorsement of any statement made
by any of our institutes, funding agencies, governments, or their representatives.

We thank the SuperKEKB team for delivering high-luminosity collisions;
the KEK cryogenics group for the efficient operation of the detector solenoid magnet;
the KEK Computer Research Center for on-site computing support; the NII for SINET6 network support;
and the raw-data centers hosted by BNL, DESY, GridKa, IN2P3, INFN, 
and the University of Victoria.

\appendix

\section{Reconstruction channels of $B$ and $D$ mesons}
\label{sec:fei_b_d_chan}

The channels used to reconstruct $B$ and $D$ mesons are
listed in Tables~\ref{tab:fei_b_chan} and \ref{tab:fei_d_chan}.
\begin{table}[htbp]
  \caption{Decay channels of $B^+$ and $B^0$ mesons used in the FEI.}
  \label{tab:fei_b_chan}
  \centering
  \begin{tabular}{@{}ll@{}} \toprule
    $B^+\to$ & $B^0\to$ \\
    \midrule
    $\bar{D}^0\pi^+$ & $D^-\pi^+$  \\
    $\bar{D}^0\pi^+\pi^+\pi^-$ & $D^-\pi^+\pi^+\pi^-$ \\
    $\bar{D}^{*0}\pi^+$ & $D^{*-}\pi^+$ \\
    $\bar{D}^{*0}\pi^+\pi^+\pi^-$ & $D^{*-}\pi^+\pi^+\pi^-$ \\
    \midrule
    $D_s^{+}\bar{D}^{0}$ & $D_s^{+}D^{-}$ \\
    $D_s^{*+}\bar{D}^{0}$ & $D_s^{*+}D^{-}$ \\
    $D_s^{+}\bar{D}^{*0}$ & $D_s^{+}D^{*-}$ \\
    $D_s^{*+}\bar{D}^{*0}$ & $D_s^{*+}D^{*-}$ \\
    \midrule
    $J/\psi \, K^+$ & $J/\psi \, \ks$ \\
    $J/\psi \, \ks\,\pi^+$ & $J/\psi \, K^+\pi^-$ \\
    $J/\psi \, K^+\pi^+\pi^-$ & \\
    \midrule
    $D^-\pi^+\pi^+$ & $D^{*-}K^+K^-\pi^+$ \\
    $D^{*-}\pi^+\pi^+$ & \\
    \bottomrule
  \end{tabular}
\end{table}
\begin{table}[htbp]
  \caption{Decay channels of $D^0$, $D^+$ and $D_s^+$ mesons used in the FEI.}
  \label{tab:fei_d_chan}
  \centering
  \begin{tabular}{@{}lll@{}} \toprule
    $D^0\to$ & $D^+\to$ & $D_s^+\to$ \\ \midrule
    $K^-\pi^+$ & $K^-\pi^+\pi^+$ & $K^+K^-\pi^+$ \\
    $K^-\pi^+\pi^0$ & $K^-\pi^+\pi^+\pi^0$ & $K^+\ks$ \\
    $K^-\pi^+\pi^+\pi^-$ & $\ks\,\pi^+$ & $K^+K^-\pi^+\pi^0$ \\
    $\ks\,\pi^+\pi^-$ & $\ks\,\pi^+\pi^0$ & $K^+\ks\,\pi^+\pi^-$ \\
    $\ks\,\pi^+\pi^-\pi^0$ & $\ks\,\pi^+\pi^+\pi^-$ & $K^-\ks\,\pi^+\pi^+$ \\
    $K^+K^-$ & $K^+K^-\pi^+$ & $K^+K^-\pi^+\pi^+\pi^-$ \\
    $K^+K^-\ks\,$ &  & $K^+\pi^+\pi^-$ \\
    & & $\pi^+\pi^+\pi^-$ \\
    \bottomrule
  \end{tabular} 
\end{table}
The sum of the corresponding branching fractions is 7.2\% for $\bp$,
6.1\% for $\bn$, 44.6\% for $D^0$, 28.6\% for $D^+$, and 17.2\% for
$D_s^+$.


\begin{thebibliography}{10}

\bibitem{Bondar:2016hva}
  A.~E.~Bondar, R.~V.~Mizuk, and M.~B.~Voloshin,
  ``Bottomonium-like states: Physics case for energy scan above the
  $B\bar{B}$ threshold at Belle~II,'' 
  Mod.\ Phys.\ Lett.\ A \textbf{32}, 1750025 (2017).

\bibitem{Meng:2007tk}
  C.~Meng and K.~T.~Chao,
  ``Scalar resonance contributions to the dipion transition rates of
  $\Upsilon(4S,5S)$ in the re-scattering model,'' 
  Phys.\ Rev.\ D \textbf{77}, 074003 (2008).

\bibitem{Simonov:2008ci}
  Y.~A.~Simonov and A.~I.~Veselov,
  ``Strong decays and dipion transitions of $\Upsilon(5S)$,''
  Phys.\ Lett.\ B \textbf{671}, 55 (2009).
  
\bibitem{Kaiser:2002bm}
  R.~Kaiser, A.~V.~Manohar, and T.~Mehen,
  ``Isospin violation in $e^+e^-\to B\bar{B}$,''
  Phys.\ Rev.\ Lett.\ \textbf{90}, 142001 (2003).
  
\bibitem{Voloshin:2012dk}
  M.~B.~Voloshin,
  ``Heavy quark spin symmetry breaking in near-threshold
  $J^{PC}=1^{--}$ quarkonium-like resonances,'' 
  Phys.\ Rev.\ D \textbf{85}, 034024 (2012).

\bibitem{Belle:2019cbt}
  R.~Mizuk \textit{et al.} [Belle Collaboration],
  ``Observation of a new structure near 10.75 GeV in the energy
  dependence of the $e^+e^-\to\Upsilon(nS)\pi^+\pi^-$ (n = 1, 2, 3) cross
  sections,'' 
  JHEP \textbf{10} (2019) 220.

\bibitem{Dong:2020tdw}
  X.~K.~Dong, X.~H.~Mo, P.~Wang, and C.~Z.~Yuan,
  ``Hadronic cross section of $e^+e^-$ annihilation at bottomonium
  energy region,'' 
  Chin.\ Phys.\ C \textbf{44}, 083001 (2020).

\bibitem{Belle:2021lzm}
  R.~Mizuk \textit{et al.} [Belle Collaboration],
  ``Measurement of the energy dependence of the $e^+e^-\to{B}\bar{B}$, 
  $B\bar{B}^*$ and $B^*\bar{B}^*$ exclusive cross sections,'' 
  JHEP \textbf{06} (2021) 137.

\bibitem{Husken:2022yik}
  N.~H\"usken, R.~E.~Mitchell, and E.~S.~Swanson,
  ``K-matrix analysis of $e^+e^-$ annihilation in the bottomonium region,''
  Phys. Rev. D \textbf{106}, 094013 (2022).
  
\bibitem{Belle-II:2024mjm}
  I.~Adachi \textit{et al.} [Belle~II Collaboration],
  ``Study of $\Upsilon(10753)$ decays to $\pi^+\pi^-\Upsilon(nS)$
  final states at Belle II,'' 
  [arXiv:2401.12021 [hep-ex]], submitted to JHEP.
  
\bibitem{Belle-II:2022xdi}
  I.~Adachi \textit{et al.} [Belle~II Collaboration],
  ``Observation of $e^+e^-\to\omega\chi_{bJ}(1P)$ and search for
  $X_b\to\omega\Upsilon(1S)$ at $\sqrt{s}$ near 10.75 GeV,''  
  Phys.\ Rev.\ Lett.\ \textbf{130}, 091902 (2023).
  
\bibitem{Belle-II:2023twj}
  I.~Adachi \textit{et al.} [Belle~II Collaboration],
 ``Search for the $e^+e^-\to\eta_b(1S)\omega$ and
  $e^+e^-\to\chi_{b0}(1P)\omega$ processes at
  $\sqrt{s}=10.745$\,GeV,''
  Phys. Rev. D \textbf{109}, 072013 (2024).
  
\bibitem{Keck:2018lcd}
  T.~Keck \textit{et al.},
  ``The Full Event Interpretation: An Exclusive Tagging Algorithm for
  the Belle~II Experiment,'' 
  Comput.\ Softw.\ Big Sci.\ \textbf{3}, 6 (2019).

\bibitem{Belle-II:2010dht}
  T.~Abe \textit{et al.} [Belle~II Collaboration],
  ``Belle~II Technical Design Report,''
  [arXiv:1011.0352 [physics.ins-det]].

\bibitem{Akai:2018mbz}
  K.~Akai \textit{et al.} [SuperKEKB],
  ``SuperKEKB Collider,''
  Nucl.\ Instrum.\ Meth.\ A \textbf{907}, 188 (2018).
  
\bibitem{Lange:2001uf}
  D.~J.~Lange,
  ``The EvtGen particle decay simulation package,''
  Nucl.\ Instrum.\ Meth.\ A \textbf{462}, 152 (2001).

\bibitem{Jadach:1999vf}
  S.~Jadach, B.~F.~L.~Ward, and Z.~Was,
  ``The Precision Monte Carlo event generator KK for two fermion final
  states in $e^+e^-$ collisions,'' 
  Comput.\ Phys.\ Commun.\ \textbf{130}, 260 (2000).

\bibitem{Sjostrand:2014zea}
  T.~Sj\"ostrand, S.~Ask, J.~R.~Christiansen, R.~Corke, N.~Desai,
  P.~Ilten, S.~Mrenna, S.~Prestel, C.~O.~Rasmussen, and P.~Z.~Skands, 
  ``An introduction to PYTHIA 8.2,''
  Comput.\ Phys.\ Commun.\ \textbf{191}, 159 (2015).

\bibitem{Barberio:1990ms}
  E.~Barberio, B.~van Eijk and Z.~Was,
  ``PHOTOS: A Universal Monte Carlo for QED radiative corrections in decays,''
  Comput.\ Phys.\ Commun.\ \textbf{66}, 115 (1991).

\bibitem{GEANT4:2002zbu}
  S.~Agostinelli \textit{et al.},
  ``GEANT4 -- a simulation toolkit,''
  Nucl.\ Instrum.\ Meth.\ A \textbf{506}, 250 (2003).

\bibitem{Kuhr:2018lps}
  T.~Kuhr \textit{et al.} [Belle~II Framework Software Group],
  ``The Belle~II Core Software,''
  Comput.\ Softw.\ Big Sci.\ \textbf{3}, 1 (2019).

\bibitem{Belle-II:2018jsg}
  E.~Kou \textit{et al.},
  ``The Belle~II Physics Book,''
  PTEP \textbf{2019}, 123C01 (2019)
  [erratum: PTEP \textbf{2020}, 029201 (2020)].
  
\bibitem{Keck:2017gsv}
  T.~Keck,
  ``FastBDT: A Speed-Optimized Multivariate Classification Algorithm
  for the Belle~II Experiment,'' 
  Comput.\ Softw.\ Big Sci.\ \textbf{1}, 2 (2017).

\bibitem{Fox:1978vu}
  G.~C.~Fox and S.~Wolfram,
  ``Observables for the Analysis of Event Shapes in $e^+e^-$ Annihilation
  and Other Processes,'' 
  Phys.\ Rev.\ Lett.\ \textbf{41}, 1581 (1978).

\bibitem{Kuraev:1985hb} 
  E.A.~Kuraev and V.S.~Fadin,
  ``On Radiative Corrections to $e^+e^-$ Single Photon Annihilation at
  High-Energy,''
  Sov.\ J.\ Nucl.\ Phys.\  {\bf 41}, 466 (1985).
  
\bibitem{Aubert:2008ab}
  B.~Aubert {\it et al.} [BaBar Collaboration],
  ``Measurement of the $e^+e^-\to b\bar{b}$ cross section
  between $\sqrt{s}$ = 10.54\,GeV and 11.20\,GeV,''
  Phys.\ Rev.\ Lett.\ {\bf 102}, 012001 (2009).

\bibitem{ParticleDataGroup:2022pth}
  R.~L.~Workman \textit{et al.} [Particle Data Group],
  ``Review of Particle Physics,''
  PTEP \textbf{2022}, 083C01 (2022) and 2023 updates.
  
\bibitem{BaBar:2008ikz}
  B.~Aubert \textit{et al.} [BaBar Collaboration],
  ``Measurement of the Mass Difference $m(B^0) - m(B^+)$,''
  Phys.\ Rev.\ D \textbf{78}, 011103 (2008).

\bibitem{Bondar:2022kxv}
  A.~E.~Bondar, A.~I.~Milstein, R.~V.~Mizuk, and S.~G.~Salnikov,
  ``Effects of isospin violation in the $e^+e^-\to B^{(*)}\bar{B}^{(*)}$ cross sections,''
  JHEP \textbf{05} (2022) 170.

\bibitem{Belle-II:2019usr}
  F.~Abudin\'en \textit{et al.} [Belle~II Collaboration],
  ``Measurement of the integrated luminosity of the Phase 2 data of
  the Belle~II experiment,'' 
  Chin.\ Phys.\ C \textbf{44}, 021001 (2020).

\bibitem{Santel:2015qga}
  D.~Santel \textit{et al.} [Belle Collaboration],
  ``Measurements of the $\Upsilon(10860)$ and $\Upsilon(11020)$
  resonances via $\sigma(e^+e^-\to \Upsilon(nS)\pi^+ \pi^-)$,'' 
  Phys.\ Rev.\ D \textbf{93}, 011101 (2016).

\bibitem{Dubynskiy:2006sg}
  S.~Dubynskiy and M.~B.~Voloshin,
  ``Possible new resonance at the $D^*\bar{D}^*$ threshold in $e^+e^-$ annihilation,''
  Mod.\ Phys.\ Lett.\ A \textbf{21}, 2779 (2006).
  
\bibitem{Ono:1985eu} 
  S.~Ono, A.~I.~Sanda and N.~A.~Tornqvist,
  ``$B$ Meson Production Between the $\Upsilon(4S)$ and $\Upsilon(6S)$
  and the Possibility of Detecting $B \bar{B}$ Mixing,''
  Phys.\ Rev.\ D {\bf 34}, 186 (1986).

\bibitem{Li:2013kya}
  X.~Li and M.~B.~Voloshin,
  ``Mixing of partial waves near $B^* \bar B^*$ threshold in $e^+e^-$
  annihilation,''
  Phys.\ Rev.\ D \textbf{87}, 094033 (2013).
  
\bibitem{Salnikov:2023cug}
  S.~G.~Salnikov, A.~E.~Bondar and A.~I.~Milstein,
  ``Coupled channels and production of near-threshold
  $B^{(*)}\bar{B}^{(*)}$ resonances in $e^+e^-$ annihilation,''
  Nucl.\ Phys.\ A \textbf{1041}, 122764, (2024).

\bibitem{Bertemes:2023LaThuile}
  M.Bertemes,
  ``Quarkonium and charm physics at Belle II,''
  QCD and High Energy Interactions (LaThuile, 2023).

\end{thebibliography}
\end{document}